\documentclass[journal]{IEEEtran} % 使用 IEEEtran 文档类，默认是双栏格式
\usepackage{caption}
\usepackage{epstopdf}
\usepackage{graphicx}
\usepackage{algorithm}
\usepackage{algorithmic}
\usepackage{booktabs}
\usepackage{makecell}
\usepackage{amsmath} %数学公式
\usepackage{amssymb}
\usepackage{multicol}
\usepackage{amsthm}
\usepackage{tikz} % 用于绘制框和标注
\usetikzlibrary{positioning} % 提供相对定位功能
\usepackage{booktabs} % 加载 booktabs 包
\usepackage{caption}   % 包含 caption 包

\nocite{*} %显示所有文献
\floatname{algorithm}{Algorithm} %算法
\renewcommand{\algorithmicrequire}{\textbf{Input:}} %输入
\usepackage{subcaption}
\renewcommand{\algorithmicensure}{\textbf{Output:}} %输出
\usepackage[colorlinks,linkcolor=black,anchorcolor=black,citecolor=black]{hyperref}

\begin{document}
\title{Task Scheduling in Space-Air-Ground Uniformly Integrated Networks with Ripple Effects}

\author{Chuan Huang, \textit{Member, IEEE}\thanks{C. Huang is with the School of Science and Engineering, the Shenzhen Future Network of Intelligence Institute, and the Guangdong Provincial Key Laboratory of Future Networks of Intelligence, The Chinese University of Hong Kong, Shenzhen 518172, China (e-mail: huangchuan@cuhk.edu.cn).}, Ran Li, \textit{Member, IEEE}\thanks{R. Li is with the Department of Information Engineering, The Chinese University of Hong Kong, Hong Kong SAR (e-mail: ranli@ie.cuhk.edu.hk).}, and Jiachen Wang\thanks{J. Wang is with the Shenzhen Future Network of Intelligence Institute, the School of Science and Engineering, and the Guangdong Provincial Key Laboratory of Future Networks of Intelligence, The Chinese University of Hong Kong, Shenzhen 518172, China (e-mail: jiachenwang1@link.cuhk.edu.cn).}}

\maketitle

\begin{abstract}
Space-air-ground uniformly integrated network (SAGUIN), which integrates the satellite, aerial, and terrestrial networks into a unified communication architecture, is a promising candidate technology for the next-generation wireless systems. Transmitting on the same frequency band, higher-layer access points (AP), e.g., satellites, provide extensive coverage; meanwhile, it may introduce significant signal propagation delays due to the relatively long distances to the ground users, which can be multiple times longer than the packet durations in task-oriented communications. This phenomena is modeled as a new ``ripple effect'', which introduces spatiotemporally correlated interferences in SAGUIN. This paper studies the task scheduling problem in SAGUIN with ripple effect, and formulates it as a Markov decision process (MDP) to jointly minimize the age of information (AoI) at users and energy consumption at APs. The obtained MDP is challenging due to high dimensionality, partial observations, and dynamic resource constraints caused by ripple effect. To address the challenges of high dimensionality, we reformulate the original problem as a Markov game, where the complexities are managed through interactive decision-making among APs. Meanwhile, to tackle partial observations and the dynamic resource constraints, we adopt a modified multi-agent proximal policy optimization (MAPPO) algorithm, where the actor network filters out irrelevant input states based on AP coverage and its dimensionality can be reduced by more than an order of magnitude. Simulation results reveal that the proposed approach outperforms the benchmarks, significantly reducing users' AoI and APs' energy consumption.
\end{abstract}


\section{Introduction}

With the rapidly growing demands for low-latency and high-speed transmissions, the upcoming sixth-generation (6G) communications are expected to support seamless global coverage \cite{6gcoverage} and various emerging applications, e.g., holographic communications, intelligent transportation, and digital twins \cite{6gapplication}. One of the promising technologies in the future 6G is the space-air-ground integrated network (SAGIN), which combines network segments across different spatial domains, i.e., space, aerial, and ground layers, to provide extensive coverage, increase network throughput, and support massive connectivity. These unique advantages make SAGIN a critical framework to overcome the limitations of conventional ground-based communication architectures, positioning it as a promising solution to meet the stringent requirements of future 6G networks \cite{6gdemands}. Numerous leading international companies and organizations, including Space Exploration Technologies Corp. (SpaceX) \cite{spacex}, Global Information Grid (GIG) \cite{gig}, OneWeb \cite{oneweb}, and International Telecommunication Union (ITU) \cite{itu}, have recognized its significance and are actively advancing its development.

In SAGIN, space layer is primarily supported by satellites, providing coverage for remote and underserved areas \cite{xiao2024space}, e.g., mountainous and rural regions; aerial layer uses unmanned aerial vehicles (UAV) \cite{kim2019multi} and high-altitude platforms (HAP) \cite{kurt2021vision} to provide highly dynamic and on-demand coverage, and substantially increase communication throughput; and ground layer, mainly implemented as cellular systems \cite{saginsegments}, supports high data-rate access for a large number of ground users. This multi-layered structural design grants SAGIN with unique communication capabilities while also introduces inherent limitations \cite{9463461}. First, each layer utilizes distinct frequency bands and different communication protocols to meet specific requirements \cite{saginsegments}, such as reliability and latency, making interoperability and efficient cross-layer coordination a significant challenge. Additionally, different spatial locations of satellites, UAVs, and ground base stations (BS) result in substantial disparities in transmission delays \cite{saginconvergence} when delivering information to ground users. These delay differences affect the timeliness of information delivery, posing critical challenges for the synchronization at the receiving ends \cite{saginchallenge}. SAGIN aims to enable the integration and coordination among these heterogeneous layers to achieve network convergence \cite{xiao2024space} and optimize the utilization of limited spectrum resources \cite{saginsegments}. However, the lack of a unified framework makes this goal challenging and contrasts with the vision of integrating the terrestrial and non-terrestrial networks under one unified framework to achieve superior performance \cite{saginconvergence}.
\begin{figure}[htbp]
    \centering
    \includegraphics[scale=0.578]{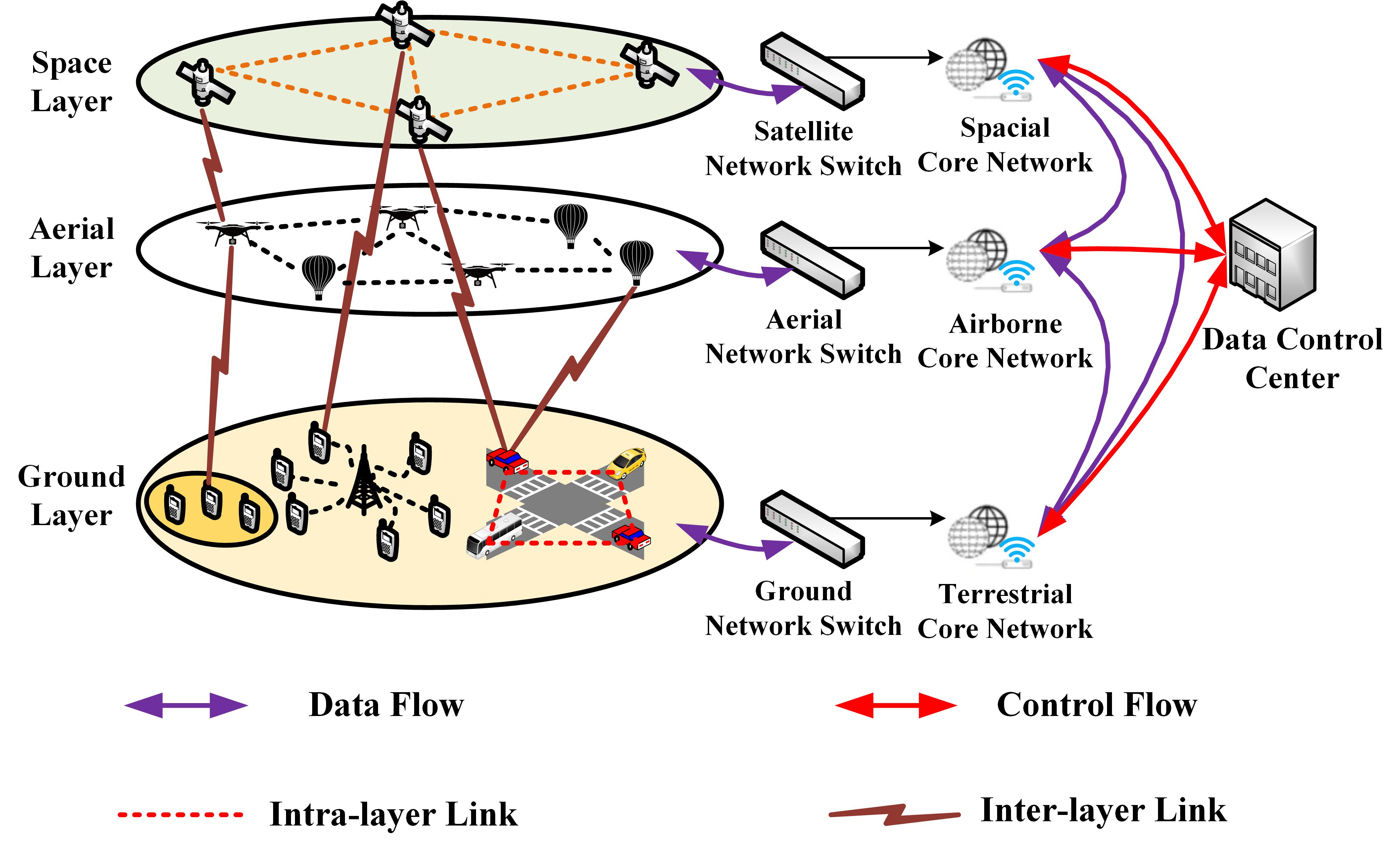}
    \captionsetup{justification=centering, font=small}  % 设置 caption 居中
    \caption{Architecture of SAGUIN.}
    \label{fig:1}       % Give a unique label
\end{figure}

To address these limitations and improve the network coordination and spectrum utilization, space-air-ground uniformly integrated network (SAGUIN) was proposed in \cite{saguin}. As illustrated in Fig. \ref{fig:1}, SAGUIN integrates the space, aerial, and ground layers under a unified communication architecture by introducing a data control center (DCC) into the conventional SAGIN framework to manage the shared spectrum resource. In this architecture, each user's data is accessed through the access point (AP) in one selected layer, transmitted via its network switch to the core network, and ultimately forwarded to the DCC for centralized processing and coordination. SAGUIN has drawn significant attentions from numerous organizations and companies: ITU, European Telecommunications Standards Institute (ETSI), and 3rd Generation Partnership Project (3GPP) have established key standards for satellite-terrestrial network convergence and proposed several application scenarios to highlight its tremendous potential in improving global connectivity, network reliability, and service scalability \cite{std1}-\cite{std10}. In industry, Huawei, SpaceX, and Amazon have successively validated technologies like satellite-to-ground direct communication \cite{satellitedirect} and the integration of the fifth-generation (5G) communication technology with satellite systems, e.g., Amazon's Kuiper project \cite{amazon}, demonstrating the feasibility and potential capabilities of SAGUIN.

Recently, significant efforts have been dedicated to addressing technical challenges and advancing key technologies in SAGUIN. The authors in \cite{hybridaccess} proposed a hybrid access method that heterogeneously combined the terrestrial and satellite networks, leveraging their complementary characteristics to maintain user-network connectivity by switching from the terrestrial to satellite networks when the terrestrial links were unavailable or congested. The authors in \cite{sdn1} introduced software-defined networking (SDN) technology, which improved the adaptability and performance of SAGUIN by enabling flexible and programmable management through the decoupling of the control plane from the data plane \cite{sdn2}. These technologies provide a solid foundation for the further development of SAGUIN \cite{wd}. Meanwhile, task-oriented communications and scheduling to enable time-sensitive tasks in SAGUIN have emerged as critical research areas \cite{saguintask1}-\cite{saguintask2}. The authors in \cite{tasktype1}-\cite{tasktype3} investigated different types of tasks in conventional SAGIN, e.g., time-sensitive tasks \cite{dove1}. Their analysis highlighted the necessity of establishing an efficient task scheduling framework to meet the demands of various tasks and optimize the resource utilization\cite{dove2}. Then, numerous scheduling strategies, including Lyapunov-based algorithms \cite{offloading} and deep reinforcement learning (DRL) \cite{taskrl1}, had been proposed to reduce energy consumption and improve spectrum efficiency \cite{taskrl2}. However, these existing strategies suffer from the curse of dimensionality and are resource-intensive when adapting to dynamic environmental changes, which limits their direct applicability to SAGUIN \cite{dove3}. To address these challenges, the authors in \cite{saguinslicing} leveraged SDN technology to implement frequency band slicing and enable dynamic resource allocation across different spatial layers.

This paper focuses on SAGUIN, where different spatial layers are integrated within one unified communication architecture to enable the efficient management of shared spectrum resources. However, when different APs transmit on the same frequency band, short-packet transmissions may suffer from unique difficulties: Higher-layer APs, e.g., satellites, provide wide-area coverage while introduce significant propagation delays due to the large distances among them and the ground users, i.e., these delays can be multiple times longer than the packet durations. As such, the interference from one particular transmitting AP may vary in the temporal domain. Furthermore, due to different spatial coverage ranges of the APs in different layers, the interferences may also vary across the spatial domain. This phenomenon is similar to the ripple effect caused by a droplet falling into water. Therefore, we use the term ‘‘ripple effect'' to describe the spatial and temporal interferences in SAGUIN and propose a new model to capture the intricate interference patterns across different layers in both the time and space domains. To evaluate its impact on the system performance, we study the task scheduling problem in SAGUIN with ripple effect and investigate the trade-off between the age of information (AoI) of users and energy consumption of APs. This trade-off arises from the fact that selecting satellites to serve users may save energy, since solar-powered satellites are modeled as consuming zero power; however, their signal transmissions may cause severe interferences to the APs and users within their coverages. The main contributions of this paper are summarized as follows:

\begin{itemize}
    \item We study the ripple effect caused by the overlap of APs' coverage areas and differences in signal propagation delays. We derive a general mathematical model to describe its temporal and spatial interfering effects on the APs and users within the coverages.
    \item With the ripple effect interference model, we formulate the task scheduling optimization problem for SAGUIN as a Markov decision process (MDP). The spatiotemporal impacts of the ripple effect result in high-dimensional state-action spaces, partial observations, and time-varying constraints, making this problem highly complex to be solved. To address high-dimensional state-action spaces, we reformulate the original problem as an equivalent Markov game, enabling each AP to infer the intentions of the others through interactive decision-making. Meanwhile, to tackle partial observations and time-varying constraints, we adopt a modified multi-agent proximal policy optimization (MAPPO) algorithm, in which the actor network filters irrelevant states using coverage characteristics and redesigns its output layer to unify resource scheduling and prevent redundant channel allocation.
\end{itemize}

The remainder of this paper is organized as follows: Section \ref{sec:model} introduces the system model and rigorously defines the ripple effect. Section \ref{sec:problem} formulates the joint task scheduling optimization problem and presents the algorithm to solve it. Section \ref{sec:results} evaluates the performance of the proposed algorithm through extensive simulations. Finally, Section \ref{sec:conclusion} concludes this paper.

\section{System Model}
\label{sec:model}

\subsection{Structure of SAGUIN}
\begin{figure}[htbp]
    \centering
    \includegraphics[scale=0.58]{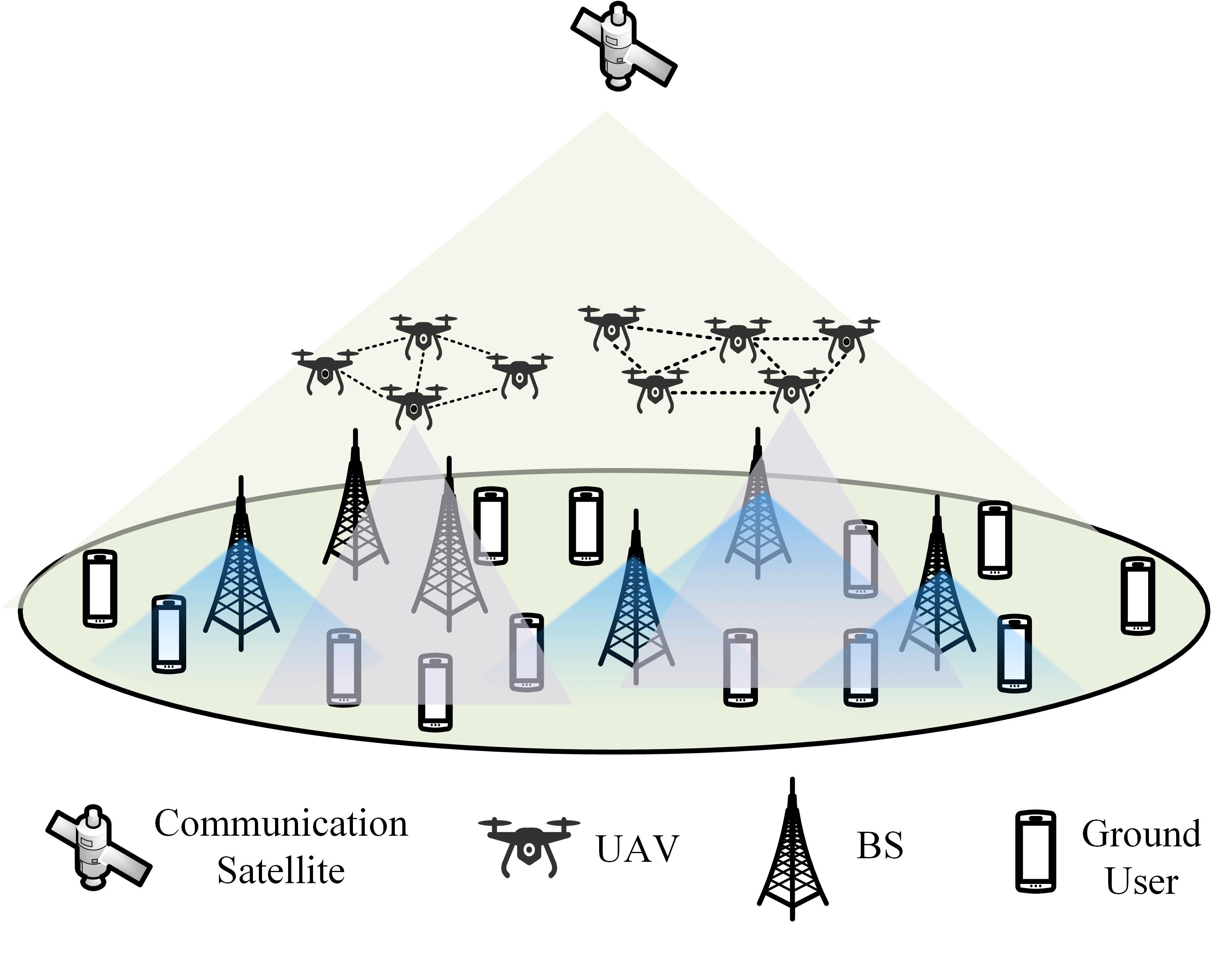}
    \captionsetup{justification=centering,font=small}
    \caption {Coverage properties of different APs in SAGUIN.}
    \label{fig:2}       % Give a unique label
\end{figure}

As illustrated in Fig. \ref{fig:2}, we consider a general SAGUIN system which consists of three communication layers: the space layer, which includes one communication satellite from a constellation network, providing extensive coverage for the target region; the aerial layer, which comprises $M$ UAVs (or HAPs) that typically follow predetermined trajectories and movement patterns to provide dynamic coverage to some specific areas in the target region; and the ground layer, which consists of $N$ stationary BSs and $U$ ground users. For indexing clarity, the satellite is labeled as $k = 0$, $M$ UAVs are indexed as $k \in \{1, 2, \ldots, M\}$, and $N$ BSs are indexed as $k \in \{M+1, M+2, \ldots, M+N\}$.
 
APs across different layers of SAGUIN exhibit distinct communication characteristics, which primarily include the following three aspects.
\begin{itemize}
    \item Coverage: The coverage relationship between AP $k$ and user $u$ depends on their geographic locations. Let $r_k$ represent the ground-level communication radius of AP $k$, and denote the coordinates of AP $k$ and user $u$ as $\mathbf{\it{X}_k} = (x_k, y_k, z_k)$ and $\mathbf{\it{X}_u} = (x_u, y_u, z_u)$, respectively. User $u$ is considered to be within the coverage of AP $k$ if the ground-level distance between them is less than or equal to $r_k$, and we set $c_{k,u} = 1$ for this case, where $c_{k,u}$ is an indication parameter; otherwise, the user is outside the coverage of the AP and we set $c_{k,u} = 0$. Hence, the coverage relationship between AP $k$ and user $u$ is expressed as
    \begin{equation}
    c_{k,u} =
    \begin{cases} 
    1, &  \sqrt{(x_u - x_k)^2 + (y_u - y_k)^2 } \leq r_k \\ 
    0, & \text{otherwise}
    \end{cases}.
    \label{eq:coverage}
    \end{equation}
    We define $\boldsymbol{c}_k \triangleq[c_{k,1}, c_{k,2}, \cdots, c_{k,U}]\in \{0, 1\}^{1 \times U}$ as the coverage vector between AP $k$ and all the $U$ users. Specifically, the satellite covers all users within the target region, i.e., $\boldsymbol{c}_0 = \boldsymbol{1}^{1 \times U}.$
    
    \item Propagation delay: The propagation delay between AP $k$ and user $u$, denoted as $d_{k,u}$, is determined by their distance and the AP types. Specifically, for BSs where $k\in\{M+1,M+2,\cdots,M+N\}$, we set $d_{k,u}=0$, since the propagation time is negligible between the BSs and their covered users. For APs of other types, the propagation delay is given as $d_{k,u}=\frac{||\mathbf{\it{X}_k}-\mathbf{\it{X}_u}||_2}{c},$ where $||\cdot||_2$ denotes the Euclidean norm and $c$ is the speed of light. To summary, $d_{k,u}$ is given by \begin{equation}
        d_{k,u} =  
    \begin{cases}  
    0, & k \!\in\! \{M\!+\!1, M\!+\!2, \cdots, M\!+\!N\} \\
    \frac{||\mathbf{\it{X}_k}-\mathbf{\it{X}_u}||_2}{c}, & \text{otherwise}
    \label{eq:delay}
    \end{cases}.
    \end{equation} Finally, the propagation delay vector for AP $k$ is defined as $\boldsymbol{d}_k \triangleq [d_{k,1}, d_{k,2}, \cdots, d_{k,U}]\in \mathbb{R}^{1 \times U}$. 

     \item Channel model: We jointly denote the path loss and the channel gain between AP $k$ and user $u$ as $g_{k,u}$. Then, the signal-to-noise ratio (SNR) of this link is given as $\gamma_{k,u} = \frac{p_{k,u} \cdot g_{k,u}}{\sigma^2}$, where $p_{k,u}$ represents the downlink transmit power and $\sigma^2$ is the noise power. The maximum data rate is calculated as $R_{k,u} = B \log_2 \left(1 + \gamma_{k,u}\right)$, where $B$ is the channel bandwidth \cite{tym}.
\end{itemize}
\subsection{Task Communications and Ripple Effect} 
To optimize the performance of SAGUIN in handling various types of tasks, we propose adopting a short-packet mechanism. In this mechanism, transmissions between APs of different types and users are mandatorily divided into short packets, with each packet being transmitted over a single channel within a fixed time duration of $\Delta t$. Apparently, by utilizing the short-packet mechanism, the occupancy status of system channels becomes strongly aligned along the time axis, ensuring maximum flexibility and utility in channel allocation. Meanwhile, to ensure the successful transmission of one complete packet within $\Delta t$, power compensation is implemented to dynamically adjust the transmission power. Specifically, we denote the transmission data size between AP $k$ and user $u$ as $S_{k,u}$, then, the transmission power is set as $p_{k,u} = \frac{(2^{\frac{S_{k,u}}{B \cdot \Delta t}} - 1) \cdot \sigma^2}{g_{k,u}}$.

\begin{figure}[htbp]
    \centering
    \includegraphics[scale=1.2]{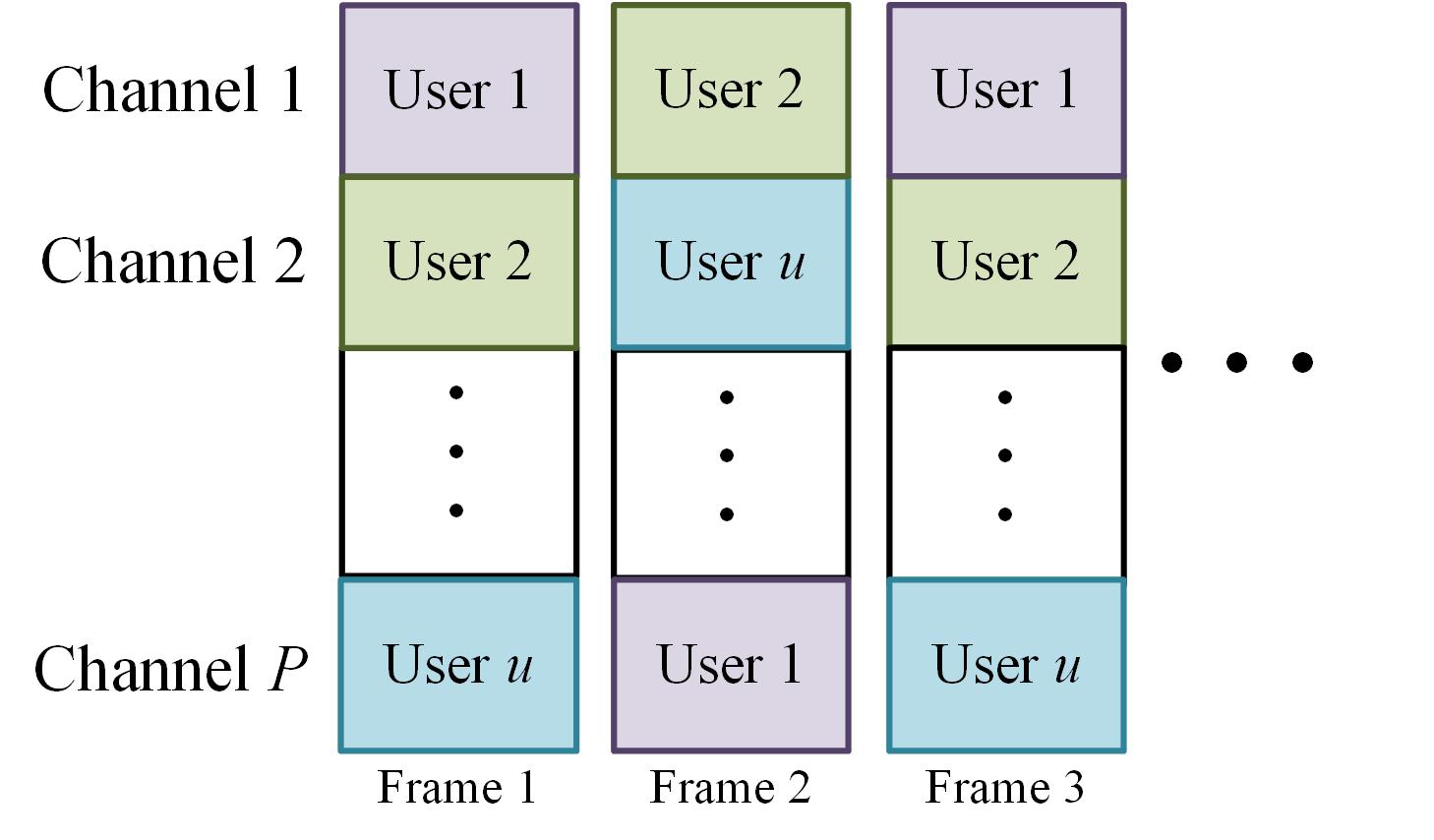}
    \captionsetup{justification=centering,font=small}  % 设置 caption 居中
    \caption{Frame structure in SAGUIN.}
    \label{fig:3}       % Give a unique label
\end{figure}

To accommodate the short-packet mechanism, we propose adopting frequency division multiple access (FDMA) in SAGUIN, where the available spectrum band is divided into $P$ orthogonal channels and the time axis is discretized into frames of fixed length $\Delta t$, as illustrated in Fig. \ref{fig:3}. The variable $a_{k,u,p}(t)$ is used to indicate the assignment of channel $p$ by AP $k$ and user $u$ at frame $t$: $a_{k,u,p}(t) = 1$ indicates that channel $p$ is assigned to AP $k$ for data transmission to user $u$ at frame $t$; and $a_{k,u,p}(t) = 0$ indicates it is not. We define a matrix $\boldsymbol{a}_k(t) \in \{0,1\}^{U \times P}$ to gather the assignment information, where the ($u,p$)-th element of $\boldsymbol{a}_k(t)$ corresponds to $a_{k,u,p}(t)$. Remarkably, each channel cannot be assigned simultaneously for data transmissions between an AP and more than one user. Thus, the channel assignment must satisfy the following constraint
\begin{equation}
\sum_{u=1}^{U} a_{k,u,p}(t)\leq 1 ,\ \forall k\in\mathcal{K},\ p\in\mathcal{P},
\label{eq:constraint1}
\end{equation} where $\mathcal{K}\triangleq\{0,1,\cdots,M+N\}$ and $\mathcal{P}\triangleq\{1,2,\cdots,P\}$ denote the sets of APs and channels, respectively.

The APs in SAGUIN have different heights and coverage areas, leading to significant variations in transmission ranges and propagation delays. For instance, higher-layer APs, such as satellites, provide the wide coverage; however, their propagation delays are often several times longer than the packet durations. After signals being transmitted by APs, they evolve dynamically over the temporal and the spatial domains, resulting in complex interference patterns. Therefore, a comprehensive analysis of the spatial and temporal properties of such interferences is crucial.
\begin{figure}[htbp]
    \centering
    \includegraphics[scale=0.65]{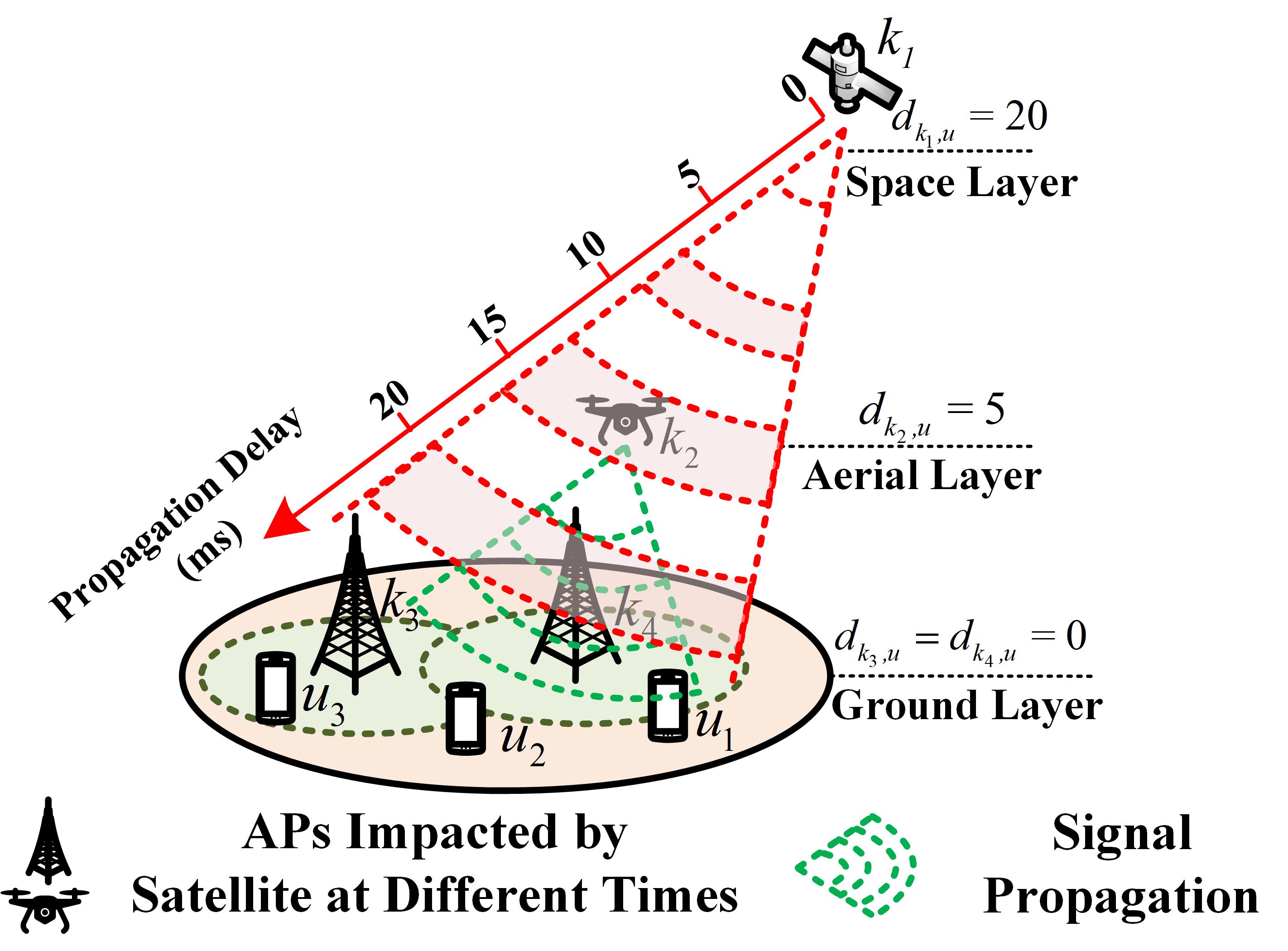}
    \captionsetup{justification=justified,singlelinecheck=false}
    \caption{Example of the spatial and temporal interferences in task-oriented communications within SAGUIN. When satellite $k_1$ transmits data over channel $p$ to its covered user $u$, the signal propagates downward and interferes with other APs that are simultaneously using the same channel $p$ to serve the same user $u$.}
    \label{fig:5}       % Give a unique label
\end{figure}
\subsubsection{Spatial Property}
The overlapping coverage areas of different APs may result in spatial interferences. When an AP transmits a signal, the intended user receives this signal after a specific propagation delay. During this process, other APs covering the same user may also transmit signals to it. Consequently, the user may simultaneously receive multiple signals from different APs on the same channel, causing significant interferences. 

\subsubsection {Temporal Property}
The heterogeneity of propagation delays across transmissions may result in temporal interferences. Specifically, when a signal is transmitted by an AP, it may collide with other signals only when it reaches the user. Therefore, all other APs must be aware of the remaining propagation time for this transmission to prevent signal interferences.

To avoid the aforementioned spatial and temporal interferences, it is sufficient to ensure that each user $u$ receives no more than one packet over each channel at any given frame $t$. Specifically, if AP $k_1$ transmits a signal to user $u$ at frame $t - d_{k_1,u}$ over channel $p$, i.e., $a_{k_1,u,p}(t - d_{k_1,u}) = 1$, user $u$ will receive this signal at frame $t$. To prevent signal interferences, any other AP $k_2$, where $k_2 \neq k_1$ and $c_{k_2,u} = 1$, must not transmit a signal to user $u$ over channel $p$ at frame $t - d_{k_2,u}$. Otherwise, this signal will also arrive at user $u$ at frame $t$, causing collisions. This constraint is mathematically formulated as 
\begin{equation}
\sum_{k \in\left\{k \in \mathcal{K} \mid c_{k, u}=1\right\}} a_{k, u, p}\left(t-d_{k, u}\right) \leq 1, \forall u \in \mathcal{U}, \ p \in \mathcal{P},
\label{eq:constraint2}
\end{equation} where $\mathcal{U}\triangleq\{1,2,\cdots,U\}$ is the users set.

\begin{figure}[htbp]
    \centering
    \includegraphics[scale=0.74]{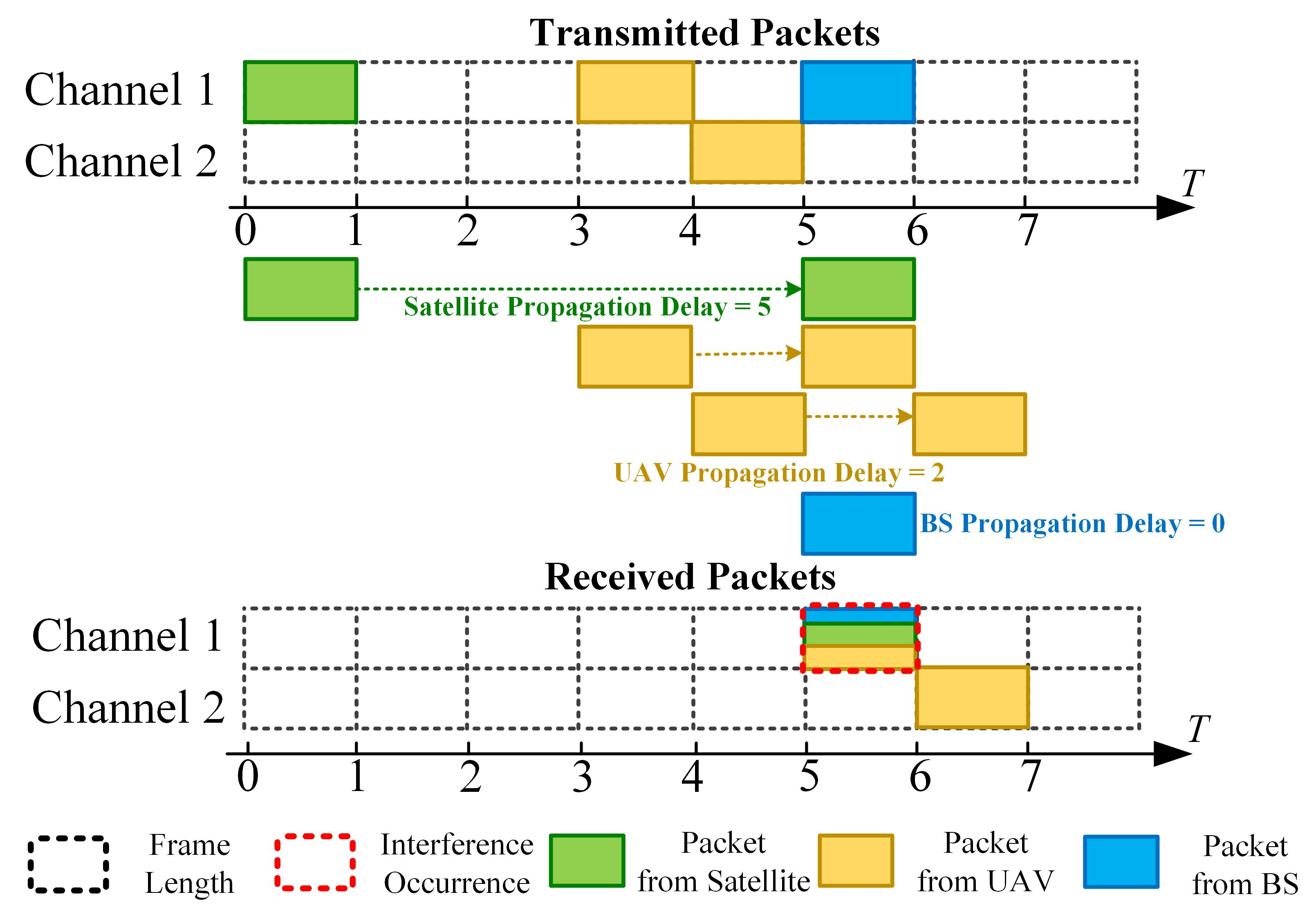}
    \captionsetup{justification=centering,font=small}  % 设置 caption 居中
    \caption{Interferences in SAGUIN.}
    \label{fig:4}       % Give a unique label
\end{figure}

Based on the above analyses, the signal interferences in time and space domains are inseparable. To be specific, interferences in SAGUIN originate at the transmitters, propagate through space, and gradually dissipate over time. To illustrate the ripple effect more concretely, we present an example in Fig. \ref{fig:4}, which involves two channels, three APs, and one user. The satellite, with a propagation delay of 5 frames, transmits a packet to the user over the first channel at $t=0$. The UAV, with a propagation delay of 2 frames, transmits another packet to the same user at $t=3$, while the BS, experiencing no propagation delay, transmits at $t=5$. All packets will simultaneously arrive at the user during the fifth frame on the first channel, resulting in significant interferences.

% The final expression for the ripple effect, as described in (\ref{eq:constraint2}), focuses on the impact of receiving ends with an emphasis on mitigating mutual interference between APs. It integrates spatial and temporal interference constraints while accurately capturing the actual arrival times of signals at user ends. This expression is composed of two components: the first term, $a_{k,u,p}(t-d_{k,u})$, checks whether AP $k$ selects channel $p$ to serve user $u$ at time $t-d_{k,u}$, and the second term, $\sum_{c_{k^\prime,u}=1}^{k^\prime:k^\prime \neq k}\sum_{u=1}^U a_{k^\prime,u,p}(t-d_{k^\prime,u})$, evaluates whether all other APs use channel $p$ to serve user $u$. This expression ensures proper coordination and prevents mutual interference among APs.
% \begin{equation}
%     \begin{aligned}
%         & a_{k,u,p} (t-d_{k,u}) 
%         + \sum_{c_{k^\prime,u}=1}^{k^\prime:k^\prime \neq k} 
%         \sum_{u=1}^U a_{k^\prime,u,p} (t-d_{k^\prime,u}) 
%         \leq 1, &\\
%     \end{aligned}
%     \label{eq:constraint2}
% \end{equation}
% $\text{for}\quad \forall u \in \mathcal{U}, \quad k, k^\prime \in \mathcal{K}, \quad \forall p \in \mathcal{P}$.

\section{Problem Formulation and Solutions}
\label{sec:problem}
This paper focuses on the typical time-sensitive tasks in SAGUIN, aiming to minimize the AoI at users and energy consumption at APs. In this section, we first formulate this problem as an MDP and analyze the difficulties to solve it. Then, we reformulate it as an equivalent Markov game and propose an MAPPO-based solution.
\subsection{MDP Formulation}
The downlink task scheduling problem in SAGUIN is formulated as an MDP, defined by following components:

\textbf{State:} The state includes two components:
\begin{itemize}
    \item AoI: For a given user $u$, its AoI at frame $t$ is defined as 
    \begin{equation}
    x_u(t)\triangleq t-T_{\text {update }},
    \label{eq:aoidef}
    \end{equation}
    where $T_{\text{update}}$ is the generation timestamp of the most recently received data packet by user $u$. Apparently, $x_u(t)$ quantifies the information freshness at user $u$. We define the AoI vector for all users as $\boldsymbol{x}(t) \triangleq[x_1(t),x_2(t), \cdots x_U(t)]$.
    \item Accumulative propagation time vector: To monitor the propagation process of each transmission between an AP $k\in\{0,1,\cdots,M\}$ and a user $u$, we define $v_{k,u,t_0}(t)$ as the real-time accumulative propagation time of the transmission starting at frame $t_0$. Specifically, when $\sum_{p=1}^P a_{k,u,p}(t)\neq 0$ and $t=t_0$, the propagation of a transmission between AP $k$ and user $u$ begins and we set $v_{k,u,t_0}(t)=1$. At each subsequent time step, the accumulative propagation time increases by 1, i.e., $v_{k,u,t_0}(t+1)=v_{k,u,t_0}(t)+1$. Once the accumulative propagation time reaches the propagation delay $d_{k,u}$, the transmitted packet arrives at the user and we set  $v_{k,u,t_0}(t)=0$. To conclude, we have \begin{equation}v_{k,u,t_0}(t\!+\!1)\! \!=\! \!\begin{cases}1, & \!\!\!\!\sum_{p=1}^P \!a_{k,u,p}(t) \!\neq\! 0,t\!=\!t_0 \\v_{k,u,t_0}(t) \!+\! 1, & \!\!\!\!1\! \leq\! v_{k,u,t_0}(t)\! < \!d_{k,u} \\0, & \!\!\!\!v_{k,u,t_0}(t) = d_{k,u}
    \label{eq:accumulative}
    \end{cases}.
    \end{equation} 
    Remarkably, at frame $t$, only the values of $v_{k,u,t_0}(t)$ for $t_0\in\{t-d_{k,u}+1,t-d_{k,u}+2,\cdots,t\}$ are useful. We define the vector $\boldsymbol{v}_{k,u}(t)$ as $\boldsymbol{v}_{k,u}(t)\triangleq[v_{k,u,t-d_{k,u}+1}(t),v_{k,u,t-d_{k,u}+2}(t),\cdots,v_{k,u,t}(t)]$ to collectively represent these values. Finally, we define the accumulative propagation time vector of AP $k$ as $\boldsymbol{v}_k(t)\triangleq[\boldsymbol{v}_{k,1}(t),\boldsymbol{v}_{k,2}(t),\cdots,\boldsymbol{v}_{k,U}(t)]\in \mathbb{N}^{1\times\sum_{u=1}^U d_{k,u}}$, where $k \in \{0, 1, 2, \ldots, M\}$.
\end{itemize}

\textbf{Action:} The action at frame $t$ is defined as $\boldsymbol{a}(t)\triangleq[\boldsymbol{a}_0(t),\boldsymbol{a}_1(t),\cdots \boldsymbol{a}_{M+N}(t)]\in\{0, 1\}^{U\times P(M+N+1)}$.

\textbf{Transition:}
The AoI for user $u$ evolves according to three distinct cases. If user $u$ receives a data packet from any BS at frame $t$, i.e., $\sum_{k=M+1}^{k=M+N}\sum_{p=1}^Pa_{k,u,p}(t)  \neq 0$, $x_u(t+1)$ is reset to 0; if user $u$ does not receive any data packet from BSs, UAVs, or satellite, $x_u(t+1)$ increases by 1, i.e., $x_u(t+1) = x_u(t) + 1$; if user $u$ does not receive any data packet from BSs but successfully receives a data packet from a UAV or satellite, $x_u(t+1)$ is updated to the smallest of the current AoI and the AoIs of the received data packets. To conclude, $x_u(t+1)$ is computed as (\ref{eq:aoi}), where $\mathcal{K}^\prime\triangleq\{k\in\{0,1,\cdots,M\}|v_{k,u,t-d_{k,u}+1}(t)=d_{k,u}-1\}$, and $\mathcal{I}(x)$ is the indicator function, which equals 1 if $x = 0$ and 0 otherwise.
\begin{figure*}[t]
\begin{equation}
x_u(t+1)=\begin{cases}
0,&\sum\limits_{k=M+1}^{k=M+N}\sum\limits_{p=1}^Pa_{k,u,p}(t)  \neq 0,\\
x_u(t)+1,&\sum\limits_{k=M+1}^{k=M+N}\sum\limits_{p=1}^Pa_{k,u,p}(t)=0 ,\\
&\sum\limits_{k=0}^M\mathcal{I}(v_{k,u,t-d_{k,u}+1}(t)-d_{k,u}+1)=0,\\
\min\{x_u(t),\min\limits_{k\in\mathcal{K}^\prime}({d_{k^\prime,u}-1})\}+1,&\sum\limits_{k=M+1}^{k=M+N}\sum\limits_{p=1}^P a_{k,u,p}(t)=0,\\
&\sum\limits_{k=0}^M\mathcal{I}(v_{k,u,t-d_{k,u}+1}(t)-d_{k,u}+1)\neq 0.
\end{cases}
\label{eq:aoi}
\end{equation}
\hrulefill
\end{figure*}

\textbf{Reward:}
Communication satellite, typically powered by solar energy, is generally self-sufficient in energy generation and consumption, so its energy consumption is not considered. Consequently, the reward function incorporates the AoI of all users and the energy consumption of APs, excluding the satellite, and is formulated as:
\begin{equation}
r(t)=-\big(\omega_1 \cdot \sum_{u=1}^{U} x_u(t) + \omega_2 \cdot\sum_{k=1}^{M+N} \sum_{u=1}^U \sum_{p=1}^P a_{k,u,p}(t) \cdot e_{k,u}\big),
\label{eq:9}
\end{equation}
where $\omega_1$ and $\omega_2$ are the importance factors for AoI and energy consumption, respectively. And $e_{k,u}$ represents the energy consumption per frame for transmission between an AP $k\in\{1,2,\cdots,M+N\}$ and a user $u$, defined as $e_{k,u} \triangleq p_{k,u} \cdot \Delta t$.

\subsection{Problem Formulation}
The average AoI for user $u$ is defined as
\begin{equation}
\bar{A}_u=\lim _{T \rightarrow \infty} \frac{1}{T} \mathbb{E}\left[\sum_{t=0}^T x_u(t)\right].
\label{eq:averageaoi}
\end{equation}
The average energy consumption of AP $k\in\{1,2,\cdots,M+N\}$ is defined as 
\begin{equation}
\bar{E}_k=\lim _{T \rightarrow \infty} \frac{1}{T} \mathbb{E}\left[\sum_{t=0}^T \sum_{u=1}^U {\sum_{p=1}^P a_{k,u,p}(t)} \cdot e_{k,u} \right].
\label{eq:averageenergy}
\end{equation}

In this paper, we aim to minimize both the average AoI at users and the energy consumption at APs, and the overall objective function is formulated as
\begin{equation}
\begin{aligned}
f =-\lim _{T \rightarrow \infty} \frac{1}{T} \mathbb{E}\left[\sum_{t=0}^T r(t)
\right]=\omega_1 \cdot \sum_{u=1}^U \bar{A}_u+\omega_2 \cdot \sum_{k=1}^{M+N} \bar{E}_k.
\end{aligned}
\label{eq:objective}
\end{equation} Then, the concerned task scheduling problem is formulated as  
\begin{equation}
\begin{aligned}
    (\mathbf{P1}) \quad \min_{\boldsymbol{a}(t)} \quad & f \\
    \text{s.t.} \quad & (\ref{eq:constraint1}), (\ref{eq:constraint2}),
\end{aligned}
\label{eq:P1}
\end{equation}
where (\ref{eq:constraint1}) limits the reuse of channels and (\ref{eq:constraint2}) specifies ripple effect.

Problem $(\mathbf{P1})$ is inherently challenging due to its high-dimensionality on state and action spaces and complex constraints in (\ref{eq:constraint1}) and (\ref{eq:constraint2}). The details are as follows.

\begin{itemize}
    \item {High dimensionality of the state and action spaces:} The dimension of the state space in ($\mathbf{P1}$) is $(M+N+1) \cdot U \cdot ( \sum_u d_{k,u}^2 + 1 )$, and the dimension of the action space is $2\cdot(M+N+1) \cdot U \cdot P$. As the number of users and channels increases, the size of the state and action spaces grows exponentially. This rapid growth significantly increases the complexity of searching for optimal solutions, making conventional MDP methods, such as dynamic programming \cite{dynamicpro} and branch and bound \cite{branch}, highly inefficient.

    \item {Constraint-related influence:} Constraints associated with the ripple effect are time-varying. Traditional DRL methods, which rely on trial-and-error mechanism, can often violate the constraints during the trial phase and thus can barely converge. 
    \item {Partial observation:} Furthermore, the considerable propagation delays of satellites and UAVs render the accumulative propagation time vectors of these APs unable to be fully obtained, resulting in partially observable state information, which restricts the algorithm's ability to achieve optimal performance.
\end{itemize}

\subsection{Markov Game Formulation}
To address these challenges, we reformulate the task scheduling problem as an equivalent Markov game, which is characterized by:

\subsubsection{Player} The players in the Markov game are all APs. Each AP observes partial state of its surrounding environment and schedules channels for data transmissions to its covered users.

\subsubsection{Action} We define $\tilde{a}_{k,p}(t)$ as the assignment status of channel $p$ to AP $k$ at frame $t$. Specifically, $\tilde{a}_{k,p}(t) = u$ indicates that channel $p$ is assigned to AP $k$ for data transmission to user $u$ at frame $t$, and $\tilde{a}_{k,p}(t) = 0$ indicates that channel $p$ is not assigned to AP $k$. We also define the action vector as $\boldsymbol{\tilde{a}_k}(t) \triangleq [\tilde{a}_{k,1}(t),\tilde{a}_{k,2}(t),\cdots,\tilde{a}_{k,P}(t)]\in \{0,1,2,\ldots,U\}^{1 \times P}$.

\subsubsection{Observation}APs of different types have distinct observations. The AoI observed by AP $k$ is defined as $\boldsymbol{y}_k(t)$, where the $u$-th element of $\boldsymbol{y}_k(t)$ is denoted as $y_{k,u}(t)$.
\begin{itemize}
     \item {Satellite:} The satellite can monitor the AoI for all users within the target area. However, due to significant uplink propagation delays, the satellite observes a delayed AoI. The satellite's observation at frame $t$ is represented as $\boldsymbol{o}_0(t) \triangleq \{\boldsymbol{y}_0(t), \boldsymbol{v}_0(t)\}$, where $\boldsymbol{y}_0(t) \triangleq \{\boldsymbol{x}(t - d_{0,u})\}$.
     \item {UAV:} Similar to the satellite, the UAV $k\in\{1,2,\cdots,M\}$ observes a delayed AoI and acquires the AoI only from the users within its coverage area. The observation of UAV $k$ at frame $t$ is represented as $\boldsymbol{o}_k(t) \triangleq \{\boldsymbol{y}_k(t), \boldsymbol{v}_k(t)\}$, where $\boldsymbol{y}_k(t) \triangleq \{x_u(t - d_{k,u}) \mid c_{k,u} = 1, u \in \mathcal{U} \}$.
     \item {BS:} The BS can instantly acquire the AoI for the users within its coverage area. Consequently, the observation of BS $k\in\{M+1,M+2,\cdots,M+N\}$ at frame $t$ is denoted as $\boldsymbol{o}_k(t)\triangleq \{\boldsymbol{y}_k(t)\}$, where $\boldsymbol{y}_k(t) \triangleq \{x_u(t) \mid c_{k,u} = 1, u \in \mathcal{U}\}$.
\end{itemize}Finally, the global observation of all APs is defined as $\boldsymbol{o}(t)\triangleq\{\boldsymbol{o}_0(t),\boldsymbol{o}_1(t),\cdots,\boldsymbol{o}_{M+N}(t)\}$.
\subsubsection{Reward} The reward of UAVs or BSs, denoted as $k\in\{1,2,\cdots, M+N\}$ is defined as
\begin{equation}
r_k(t) = -\bigg( \omega_1\cdot\sum_{u=1}^{U}y_{k,u}(t) +\omega_2 \cdot\sum_{p=1}^{P}  (1-\mathcal{I}(\tilde{a}_{k,p}(t)))\cdot e_{k,u}\bigg).
\label{eq:gamereward}
\end{equation}The first term penalizes higher AoI, and the second term accounts for the energy consumption of AP $k$. Notably, for satellite, the reward $r_0(t) = -( \sum_{u=1}^{U} y_{0,u}(t)).$

\subsubsection{Transition}The observation transition from $\boldsymbol{o}_k(t)$ to $\boldsymbol{o}_k(t+1)$ can be divided into two parts: the transition rule for $\boldsymbol{v}_k(t)$ follows (\ref{eq:accumulative}), and the transition rule for $\boldsymbol{y}_k(t)$ follows (\ref{eq:aoi}).

\textit{Remark 1:} The Markov game formulation addresses the challenges of high dimensionality through interactive decision-making among multiple players. However, despite mitigating the dimensionality issue, some challenges remain unresolved.
\begin{itemize}
    \item {Constraint-related influence}: Traditional multi-agent deep reinforcement learning (MADRL) methods for solving Markov games still rely on a trial-and-error mechanism and thus cannot effectively handle the time-varying constraints introduced by the ripple effect.
    
    \item {Lack of global information}: In practice, APs operate under partial observability and lack access to global state information or the policies of other APs.
\end{itemize}
\begin{figure*}[!h]
    \centering
	\includegraphics[scale=0.85]{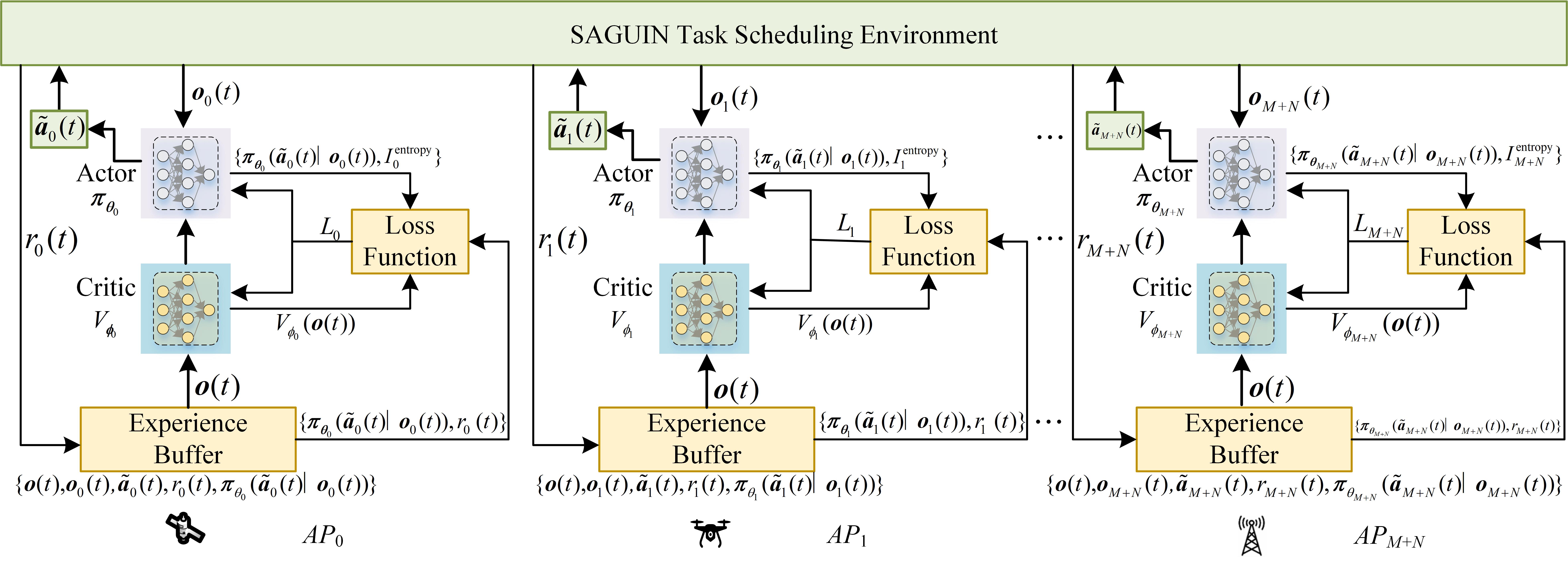}
	\caption{Structure of proposed MAPPO-based algorithm.}
	\label{fig:6}       % Give a unique label
\end{figure*}
\subsection{Structure of Proposed MAPPO-based Algorithm}
In this part, we propose an MAPPO-based algorithm to solve the formulated Markov game. The MAPPO framework employs $M+N+1$ independent PPO instances, with each instance dedicated to controlling the actions of a specific AP. Each PPO instance comprises three core components: an actor network $\boldsymbol{\pi}_{\boldsymbol{\theta}_k}$, a critic network $V_{\boldsymbol{\phi}_k}$, and an experience buffer $D_k$.

\subsubsection{Actor Network}
The actor network of PPO $k$ is a fully connected neural network that takes the local observation $\boldsymbol{o}_k(t)$ as input. It consists of two hidden layers, each with 64 neurons and Tanh activation function, followed by an output layer with a Softmax activation function. The network outputs a set of probability distributions for each of the $P$ channels. For each channel $p$, the network produces a Softmax distribution $\boldsymbol{\pi}_{\boldsymbol{\theta}_k}(\cdot \mid \boldsymbol{o}_k(t))$ over $U+1$ possible actions: servicing one of $U$ users or not providing service. At each time step $t$, the action $\tilde{a}_{k,p}(t)$ for channel $p$ is independently sampled from its corresponding Softmax probability distribution. Finally, the action vector $\boldsymbol{\tilde{a}}_k(t)$ is formed by combining the actions of all channels $\{\tilde{a}_{k,p}(t)\}_{p=1}^P$.

\subsubsection{Critic Network}
The critic network shares a similar structure with the actor network, featuring two hidden layers but differing in its input and output. It takes global observation $\boldsymbol{o}(t)$ as input and processes them through the hidden layer with 64 neurons and Tanh activation. The network outputs a single scalar value, which represents an estimate of the expected cumulative discounted rewards of the current state.

\subsubsection{Experience Buffer}
The experience buffer of PPO $k$ stores the generated experiences during offline training phase, represented as $\left\{\boldsymbol{o}(t),\boldsymbol{o}_k(t), \boldsymbol{\tilde{a}}_k(t), r_k(t), \boldsymbol{\pi}_{\boldsymbol{\theta}_k}(\boldsymbol{\tilde{a}}_k(t) \mid \boldsymbol{o}_k(t))\right\}$.

\subsection{Offline Training and Online Deployment} 
\subsubsection{Offline Training} 
In the offline training phase, we first simulate an offline SAGUIN environment based on historical observed samples, including state evolution and reward generation. Then, we conduct offline training process by interacting with this simulated environment.

\textbf{Offline Environment Simulation:} 
\begin{itemize}
    \item State Evolution:
    Based on current action $\{\boldsymbol{\tilde{a}}_k(t)\}_{k=0}^{M+N}$, the environment transitions to the next state, which is observed as $\boldsymbol{o}(t+1)$. Each AP $k$ receives its local observation $\boldsymbol{o}_k(t+1)$, derived by applying (\ref{eq:accumulative}) and (\ref{eq:aoi}).
    \item Reward Generation: Each AP $k$ receives reward $r_k(t)$ calculated according to (\ref{eq:gamereward}).
\end{itemize}

\textbf{Offline Training Process:} As illustrated in Fig. \ref{fig:6}, we alternately generate experiences by deploying the latest actor-critic network and update the actor-critic network according to the latest generated experiences. We specify these two steps as follows:
\begin{itemize}
    \item {Generation of Experiences:} 
    During this step, $M+N+1$ modified PPOs interact with the offline environment to sequentially generate $N_B$ experiences, where $N_B$ is the size of the experience buffer. Specifically, at the $t$-th frame, we send AP observations $\{\boldsymbol{o}_k(t)\}_{k=0}^{M+N}$ to actor networks and get actions $\{\boldsymbol{\tilde{a}}_k(t)\}_{k=0}^{M+N}$ and probabilities $\{\boldsymbol{\pi}_{\boldsymbol{\theta}_k}(\boldsymbol{\tilde{a}}_k(t) \mid \boldsymbol{o}_k(t))\}_{k=0}^{M+N}$. Next, we send $\{\boldsymbol{\tilde{a}}_k(t)\}_{k=0}^{M+N}$ to the simulated offline environment to obtain rewards $\{r_k(t)\}_{k=0}^{M+N}$ and the next set of AP observations $\{\boldsymbol{o}_k(t+1)\}_{k=0}^{M+N}$. We repeat the above procedures for $N_B$ iterations from the $t$-th to the $(t + N_B - 1)$-th frame and pack the related information, i.e., $\left\{\left(\boldsymbol{o}(t), \boldsymbol{o}_k(t), \boldsymbol{\tilde{a}}_k(t), r_k(t), \boldsymbol{\pi}_{\boldsymbol{\theta}_k}(\boldsymbol{\tilde{a}}_k(t) \mid \boldsymbol{o}_k(t))\right)\right\}_t^{t+N_B-1}$ into PPO $k$'s experience buffer $D_k$.

    \item {Update of Actor-Critic Networks:} The training process for the actor-critic networks involves performing $Q$ epochs of optimization on the collected experiences. At the start of each epoch, the actor and critic networks for PPO $k$ are denoted as $\boldsymbol{\pi}_{\boldsymbol{\theta}'_k}$ and $V_{\boldsymbol{\phi}'_k}$, respectively. The critic network is optimized by minimizing the mean squared error between the predicted value and the target value. The loss function for the critic is defined as \begin{equation} 
    L_k^{\text {critic }}(\boldsymbol{\phi}_k)=\frac{1}{2}[V_{\boldsymbol{\phi}_k}(\boldsymbol{o}(t))-R_k(t)]^2,
    \label{eq:criticloss}
    \end{equation} where $V_{\boldsymbol{\phi}_k}(\boldsymbol{o}(t))$ is the scalar output of the critic network, and $R_k(t)=\sum_{i=0}^{N_B-1} \gamma^i r_k(t+i)$ represents the discounted cumulative reward, with $\gamma$ as the discount factor. The actor network aims to maximize the expected cumulative reward under its policy $\boldsymbol{\pi}_k$. The optimized objective function is defined as
\begin{equation}
\begin{aligned}
J(\boldsymbol{\theta}_k) &= \mathbb{E} \Big[ \min \Big( \frac{\boldsymbol{\pi}_{\boldsymbol{\theta}_k}(\boldsymbol{\tilde{a}}_k(t)|\boldsymbol{o}_k(t))}{\boldsymbol{\pi}_{\boldsymbol{\theta}'_k}(\boldsymbol{\tilde{a}}_k(t)|\boldsymbol{o}_k(t))} A, \\
&\phantom{=} \text{clip}\Big(\frac{\boldsymbol{\pi}_{\boldsymbol{\theta}_k}(\boldsymbol{\tilde{a}}_k(t)|\boldsymbol{o}_k(t))}{\boldsymbol{\pi}_{\boldsymbol{\theta}'_k}(\boldsymbol{\tilde{a}}_k(t)|\boldsymbol{o}_k(t))}, 1-\epsilon, 1+\epsilon \Big) A \Big) \Big],
\end{aligned}
\end{equation} which is approximated by averaging the stored experiences in $D_k$ during the training process. The terms $\boldsymbol{\pi}_{\boldsymbol{\theta}_k}$ and $\boldsymbol{\pi}_{\boldsymbol{\theta}'_k}$ represent the current policy and the old policy, respectively. The advantage function is defined as $A = r_k(t)+\gamma V_{\boldsymbol{\phi}_k}(\boldsymbol{o}(t+1)) - V_{\boldsymbol{\phi}_k}(\boldsymbol{o}(t))$. The clipping function is given by $\text{clip}(r, 1-\epsilon, 1+\epsilon) = \max(1-\epsilon, \min(r, 1+\epsilon)),$ where $\epsilon$ is a constant. The loss function for the actor network $\boldsymbol{\pi}_{\boldsymbol{\theta}_k}$ is given by \begin{equation} 
    L_k^{\text {actor }}({\boldsymbol{\theta}_k})=-J(\boldsymbol{\theta}_k).
    \label{eq:actorloss}
    \end{equation} The total loss function for PPO $k$ is written as
    \begin{equation}
L_k = L_k^{\text{actor}} + L_k^{\text{critic}} - \beta \cdot I_k^{\text{entropy}}, 
\label{eq:total loss}
\end{equation} where the entropy regularization term is  $I_k^{\text{entropy}} = \mathbb{E} \left[ -\sum \boldsymbol{\pi}_{\boldsymbol{\theta}_k}(\boldsymbol{\tilde{a}}_k(t) | \boldsymbol{o}_k(t)) \log \boldsymbol{\pi}_{\boldsymbol{\theta}_k}(\boldsymbol{\tilde{a}}_k(t) | \boldsymbol{o}_k(t)) \right]$ and $\beta$ is its corresponding weight, typically a small positive value. Using the backpropagation method, both networks of PPO $k$ are updated based on (\ref{eq:total loss}) \cite{adaptive}.
\end{itemize}

The details of the offline training algorithm are summarized in Algorithm \ref{al:1}.
\begin{algorithm}[htbp]

    \caption{MAPPO-based Task Scheduling Optimization Algorithm for SAGUIN \label{al:1}} 

    \renewcommand{\algorithmicrequire}{\textbf{Input:}}
    \renewcommand{\algorithmicensure}{\textbf{Output:}}
    \renewcommand{\algorithmicfor}{\textbf{For}}
    \renewcommand{\algorithmicendfor}{\textbf{End for}}
    \renewcommand{\algorithmicreturn}{\textbf{Return}}

    \begin{algorithmic}[1]

        \REQUIRE Experience replay buffer ${D}_k$, total number of episodes ${E}$, episode length ${T}$, number of policy update steps $Q$, replay buffer size $N_B$.
        \ENSURE Actor network parameters $\{\boldsymbol{\theta}_k\}_{k=0}^{M+N}$ and critic network parameters $\{\boldsymbol{\phi}_k\}_{k=0}^{M+N}$.

    \FOR{AP $k=0$ to $M+N$}
        \STATE Initialize actor network parameter $\boldsymbol{\theta}_k$, critic network parameter $\boldsymbol{\phi}_k$, and replay buffer $D_k$;
    \ENDFOR
   
    \FOR{each episode $e =1$ to $E$}
        \STATE Initialize AoI vector $\boldsymbol{x}(t)$ as $\boldsymbol{0}^{1\times U}$ and coverage vector $\{\boldsymbol{c}_k\}_{k=0}^{M+N}$;
        \STATE Reset environment and
         get initial global observation;
        \FOR{each time step $t =1$ to $T$}
            \STATE Send local observation set $\{\boldsymbol{o}_k(t)\}_{k=0}^{M+N}$ to $M+N+1$ modified PPOs and derive probability distributions $\{\boldsymbol{\pi}_{\boldsymbol{\theta}_k}(\cdot| \boldsymbol{o}_k(t))\}_{k=0}^{M+N}$;
            \STATE Sample action $\{\boldsymbol{\tilde{a}}_k(t)\}_{k=0}^{M+N}$ from these probability distributions and get $\{\boldsymbol{\pi}_{\boldsymbol{\theta}_k}(\boldsymbol{\tilde{a}}_k| \boldsymbol{o}_k(t))\}_{k=0}^{M+N}$;
            \STATE Send $\{\boldsymbol{\tilde{a}}_k(t)\}_{k=0}^{M+N}$ to the simulated offline environment and derive $\{r_k(t)\}_{k=0}^{M+N}$ and $\{\boldsymbol{o}_k(t+1)\}_{k=0}^{M+N}$ based on (\ref{eq:gamereward}), (\ref{eq:accumulative}), and (\ref{eq:aoi});
            \STATE Store the obtained experience tuple during training $\{ \boldsymbol{o}(t),\boldsymbol{o}_k(t), \boldsymbol{\tilde{a}}_k(t), r_k(t), \boldsymbol{\pi}_{\boldsymbol{\theta}_k}(\boldsymbol{\tilde{a}}_k| \boldsymbol{o}_k(t))\}_{t}^{t+N_B-1}$ in experience buffer $D_k$;
        \ENDFOR
        
        \FOR{update iteration $q=1$ to $Q$}
            \FOR{AP $k=0$ to $M+N$}
                \STATE Load the stored experiences from $D_k$;
                \STATE Calculate the loss $L_k$ based on (\ref{eq:total loss});
                \STATE Update the actor network parameter $\boldsymbol{\theta}_k$ and critic network parameter $\boldsymbol{\phi}_k$ by back propagating $L_k$;
            \ENDFOR
        \ENDFOR
        \STATE Empty all the experience buffers;
    \ENDFOR
    \RETURN Trained network parameters $\{\boldsymbol{\theta}_k\}_{k=0}^{M+N}$ and $\{\boldsymbol{\phi}_k\}_{k=0}^{M+N}$.
    \end{algorithmic}
\end{algorithm}

\subsubsection{Online Application}
The online application phase closely resembles the offline training process. However, the key difference is that during online phase, only trained actor networks are used to interact with the real SAGUIN environment. Furthermore, the AoI vector $\boldsymbol{x}(t)$ is dynamically updated based on real-time interactions with the real environment.

\section{Simulation Results and Discussions}
\label{sec:results}
In this section, we evaluate the effectiveness and robustness of the proposed MAPPO-based algorithm. First, we specify the simulation settings. Then, we present experimental results under various scenarios to assess the performance of the proposed algorithm.
\begin{figure*}[htbp]
    \centering
	\includegraphics[scale=0.24]{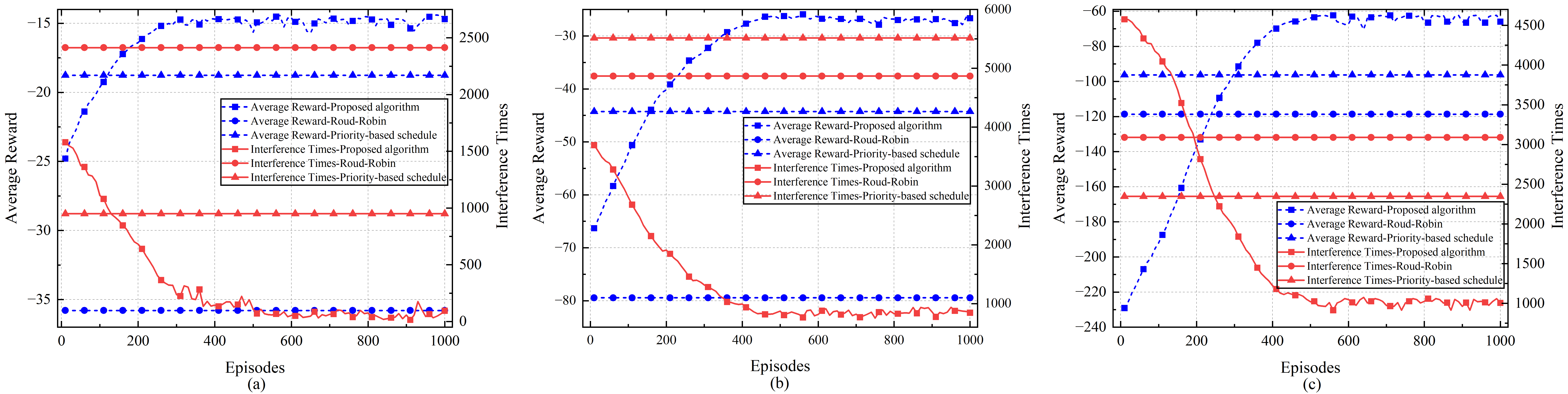}
    \captionsetup{justification=justified, font=small}
	\caption{Comparison of average reward  and interference times across different algorithms. (a) Small-scale scenario with 1 satellite, 1 UAV, 1 BS, 3 channels, and 5 users; (b) Medium-scale scenario with 1 satellite, 2 UAVs, 3 BSs, 5 channels, and 8 users; (c) Large-scale scenario with 1 satellite, 3 UAVs, 4 BSs, 10 channels, and 10 users.}
	\label{fig:ripple}       % Give a unique label
\end{figure*}
\begin{figure*}[htbp]
    \centering
	\includegraphics[scale=0.245]{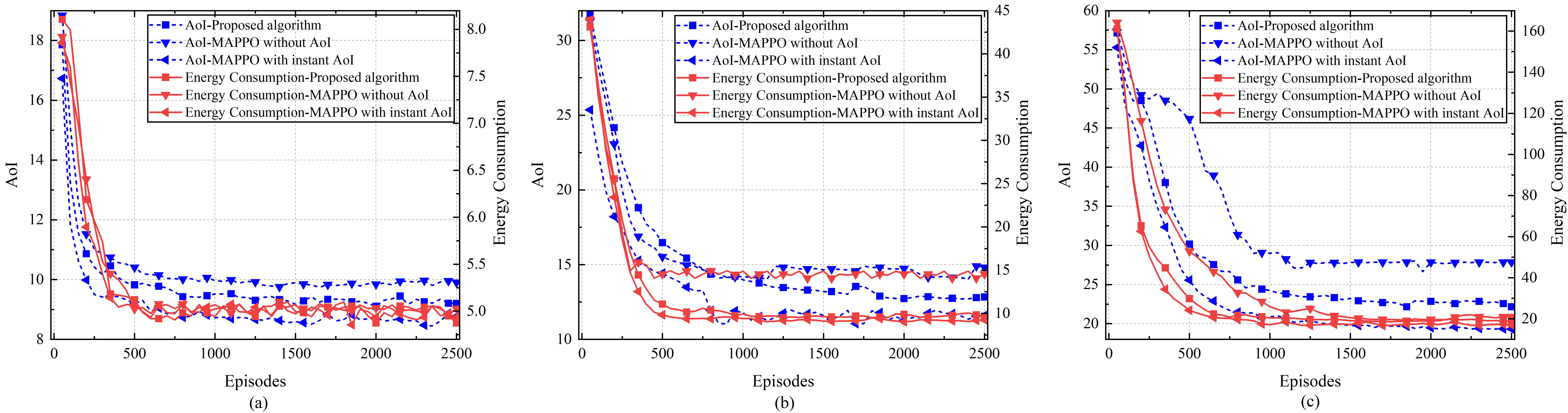}
    \captionsetup{justification=justified, font=small}
	\caption{Performance comparison across different MAPPO-based algorithms. (a) Small-scale scenario with 1 satellite, 1 UAV, 1 BS, 3 channels, and 5 users; (b) Medium-scale scenario with 1 satellite, 2 UAVs, 2 BSs, 5 channels, and 6 users; (c) Large-scale scenario with 1 satellite, 3 UAVs, 4 BSs, 6 channels, and 10 users.}
	\label{fig:7}       % Give a unique label
\end{figure*}
\subsection{Simulation Settings}

\subsubsection{Hyperparameters} The discount factor $\gamma$ is set to 0.95 to emphasize future rewards; the importance factor $\omega_1$ and $\omega_2$ are set to 0.5; the clip parameter for MAPPO is set to 0.2 to prevent overly large policy updates; actor and critic networks are trained with a learning rate of 0.001 to ensure steady optimization; each agent performs 50 epochs of policy updates per iteration, which is carefully chosen to strike a balance between sufficient policy optimization and manageable computational overhead.

\subsubsection{Propagation Delay}
The frame length $\Delta t$ is set to 1 millisecond (ms), and the propagation delay for all users covered by the satellite is set to 20 ms. Furthermore, we select representative propagation delay values of 1 ms and 5 ms to model the propagation delay between UAVs and users. The propagation delays for BSs are ignored.

\subsubsection{Energy Consumption}
\begin{itemize}
    \item Space layer: The energy consumption for user $u$ served by the satellite is set to 0, i.e., $e_{0,u}=0$, for all $u\in \mathcal{U}$.
    \item Aerial layer: The channel gain between UAV $k$ and user $u$ is modeled using the free-space path loss model, i.e., $g_{k,u} = (\frac{c}{4\pi q_{k,u}})^2$, where $q_{k,u}$ represents the distance between UAV $k$ and user $u$, and $c$ is the speed of light. The energy consumption is computed as $e_{k,u}=p_{k,u} \cdot \Delta t$, where $p_{k,u} = \frac{(2^{\frac{S_{k,u}}{B \cdot \Delta t}} - 1) \cdot \sigma^2}{g_{k,u}}$.
    \item Ground layer: The channel gain between BSs and users is modeled using COST-231 Hata model and is given by $g_{k,u} = 128.1 + 37.6 \log_{10}q_{k,u}$ (dB). The calculation of energy consumption for BSs is the same as that for UAVs.
\end{itemize}

\subsubsection{Benchmarks}To evaluate the performance of the proposed algorithm, we compare it with the following benchmarks:
\begin{itemize}
\item Round-Robin: In this mechanism, the APs serve the users within their coverage areas in a cyclic manner.

\item Priority-based schedule: In this mechanism, the priority of each user is determined based on their current AoI. Users with higher AoI values are considered to have more outdated information and are therefore prioritized for service.
\item MAPPO without AoI: This method adopts the proposed MAPPO-based algorithm. However, the AoI is not fed back to satellite and UAVs. In other words, the observations available to these APs only consist of accumulative propagation time vectors.
\item MAPPO with instant AoI: This method also employs our proposed MAPPO-based algorithm, with AoI immediately fed back to all APs. Apparently, this method is impractical for real-world implementation and is only used to provide an optimal performance bound for our proposed algorithm.
\item AoI lower bound: This benchmark represents the theoretical minimum of the average AoI achievable under ideal conditions. The detailed derivation of this lower bound is provided in the Appendix.
\end{itemize}

\subsection{Numerical Results}
In Fig. \ref{fig:ripple}, the average reward and interference times of different algorithms are compared across small-, medium-, and large-scale network scenarios. The metric interference times refers to the number of occurrences in which a user receives packets from more than one AP over the same channel within the same frame during an episode. This metric is averaged over every 10 episodes (i.e., 10,000 frames). In all scenarios, the proposed algorithm demonstrates stable and consistent learning behavior. The average reward increases steadily during the initial training phase and eventually converges to a value significantly higher than those achieved by the benchmarks. Simultaneously, interference times shows a sharp decline in the early training stages and stabilizes at a much lower level compared to the benchmarks. Notably, in the small-scale scenario, interferences are almost entirely mitigated. These results clearly demonstrate the effectiveness of the proposed algorithm in identifying and mitigating spatiotemporal interferences, as well as its capability to learn interference-aware scheduling policies within SAGUIN.

In Fig. \ref{fig:7}, we compare the performance of various MAPPO algorithms across different scales in SAGUIN. In all scenarios, MAPPO with delayed AoI, i.e., the proposed algorithm, consistently outperforms the one without AoI information. Specifically, in the small-scale scenario, the average AoI is decreased by 10\%. In the medium-scale scenario, both the average AoI and energy consumption are reduced by approximately 16\%. The improvements are even more pronounced in the large-scale scenario, with a 20\% decrease in average AoI and a 5\% reduction in energy consumption. Notably, the performance of the proposed algorithm closely approaches that of MAPPO with instant AoI, exhibiting less than a 5\% degradation in both metrics across all scenarios. These results highlight the practicality and effectiveness of using delayed AoI as an alternative in real-world deployments, particularly in environments where real-time information is unavailable.
\begin{figure*}[htbp]
    \centering
	\captionsetup{justification=justified, font=small}
    \includegraphics[scale=0.3]{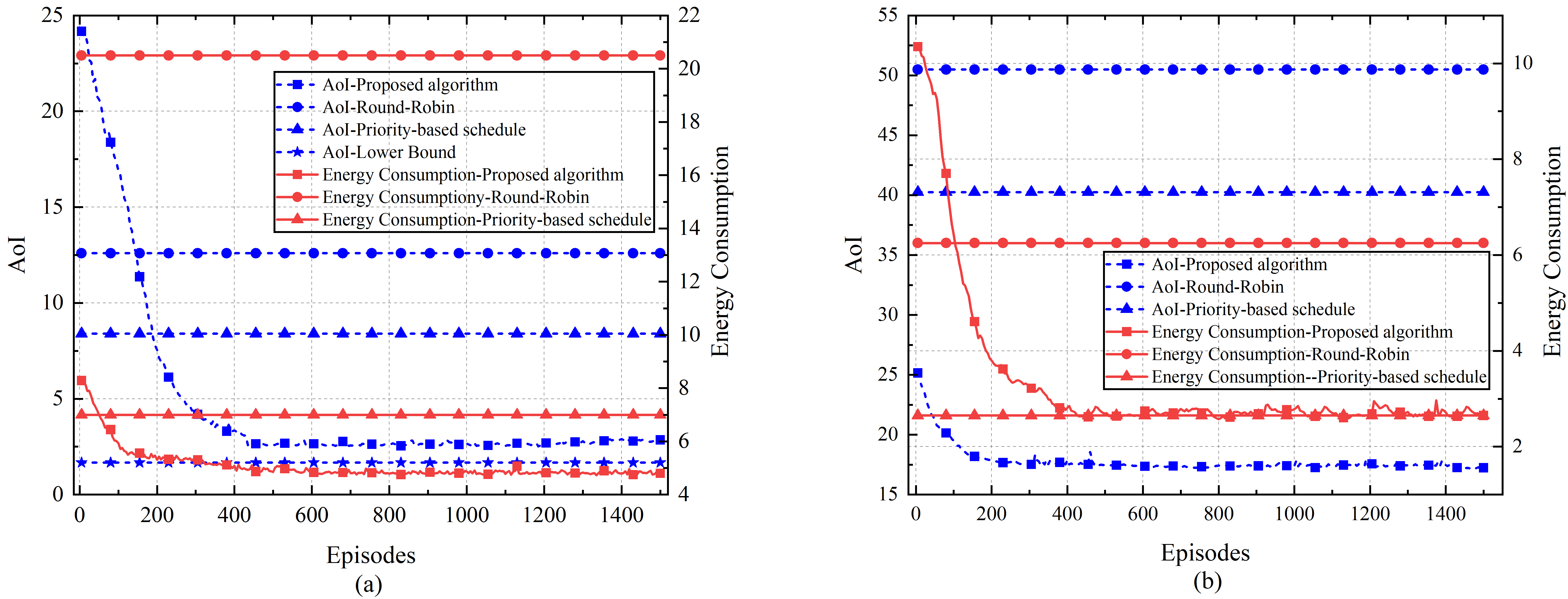}
	\caption{Performance comparison across different coverage conditions. (a) Full coverage scenario where every AP covers all users; (b) Partial coverage scenario where the UAV only covers 3 users and the BS covers 2 users. Both (a) and (b) correspond to the small-scale scenario with 1 satellite, 1 UAV, 1 BS, 3 channels, and 5 users.}
	\label{fig:8}       % Give a unique label
\end{figure*}

\begin{figure*}[htbp]
    \centering
	\captionsetup{justification=justified, font=small}
    \includegraphics[scale=0.3]{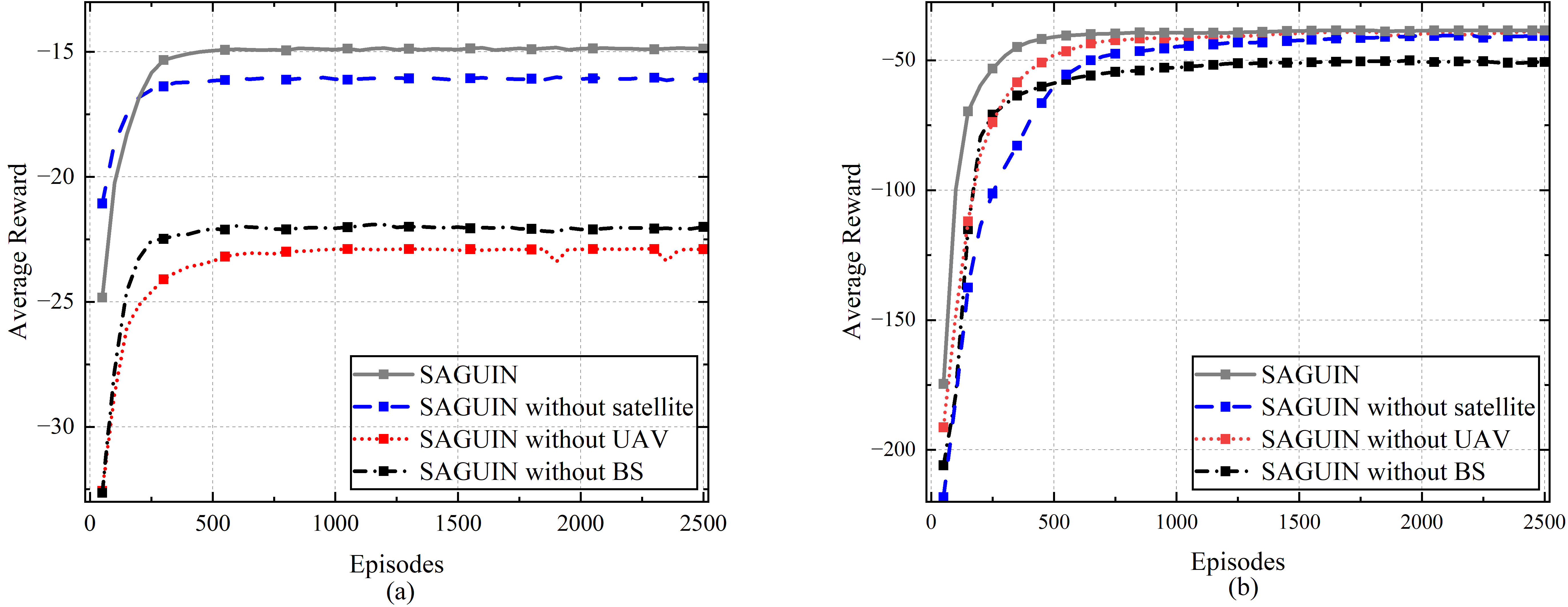}
	\caption{Average reward under different network configurations. (a) Small-scale scenario with 1 satellite, 1 UAV, 1 BS, 3 channels, and 5 users; (b) Large-scale scenario with 1 satellite, 3 UAVs, 4 BSs, 6 channels, and 8 users.}
	\label{fig:9}       % Give a unique label
\end{figure*}
In Fig. \ref{fig:8}, the performance of various algorithms is evaluated under different AP coverage conditions. Fig. \ref{fig:8}(a) simulates the full coverage scenario, where each AP covers all users. In this setting, the proposed algorithm significantly outperforms both benchmarks in terms of AoI and energy consumption, achieving an average AoI that closely approaches the theoretical lower bound. Fig. \ref{fig:8}(b) illustrates the partial coverage scenario, where the UAV covers only 3 users and the BS covers 2 users. It is observed that the proposed algorithm achieves an average AoI approximately 60\% lower than that of the benchmarks, while maintaining an energy consumption level comparable to the priority-based schedule. These results underscore the superior efficiency, robustness, and adaptability of the proposed algorithm across heterogeneous AP coverage conditions in SAGUIN.

In Fig. \ref{fig:9}, we evaluate the performance of the proposed algorithm under various network configurations, including the complete SAGUIN architecture, as well as ablated versions without the satellite, UAV, or BS. Fig. \ref{fig:9}(a) simulates a small-scale scenario where each layer contains only one AP. It is observed that removing any AP leads to a significant degradation in the average reward, highlighting the importance of cross-layer collaboration in effectively serving users and maintaining system capacity. We have also simulated a large-scale scenario in Fig. \ref{fig:9}(b) with multiple APs in each layer. In this case, the impact of removing individual APs is less significant. This is because the large number of APs in SAGUIN leads to overcapacity, which diminishes the performance gains from integrating the space, aerial, and ground layers. In conclusion, SAGUIN offers more integration gains in smaller-scale networks, and our proposed algorithm effectively helps to achieve these gains.

\begin{figure*}[htbp]
    \centering
	\captionsetup{justification=justified, font=small}
    \includegraphics[scale=0.26]{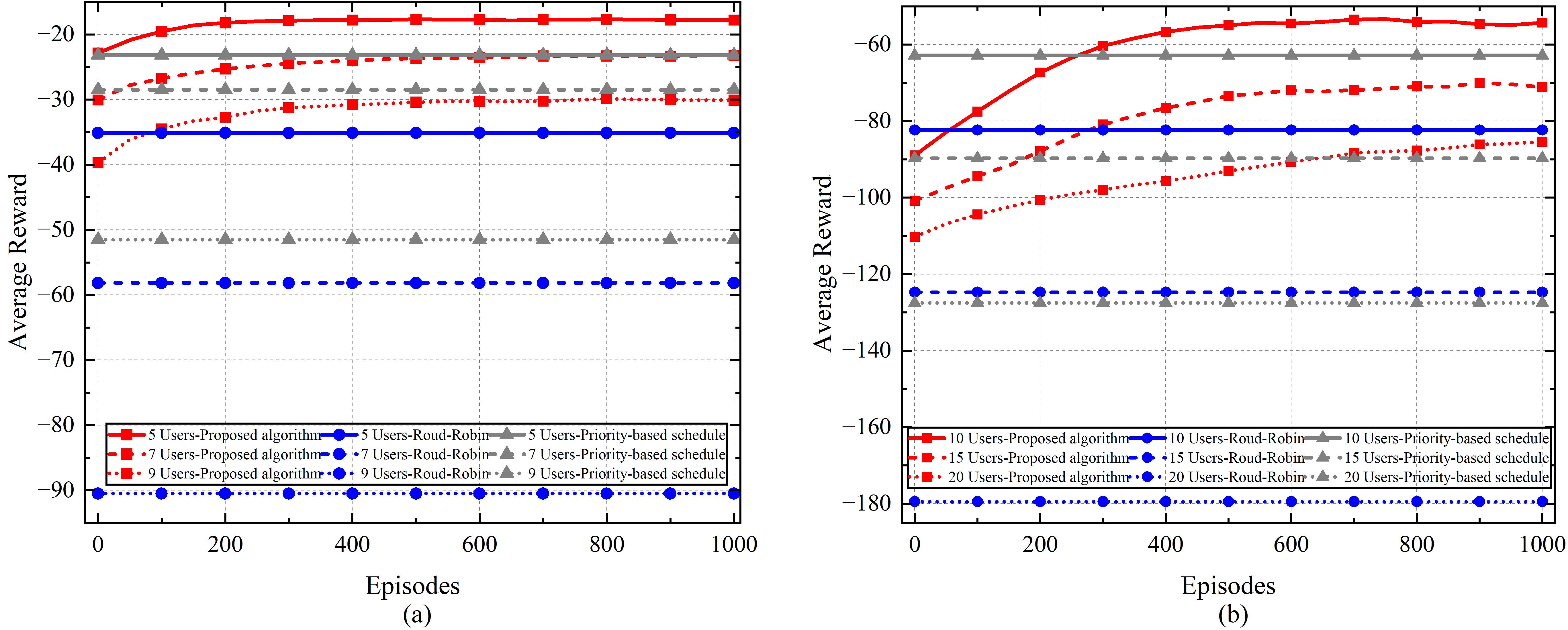}
	\caption{Performance comparison across different algorithms with varying numbers of users. (a) Small-scale scenario with 1 satellite, 1 UAV, 1 BS, and 3 channels; (b) Medium-scale scenario with 1 satellite, 2 UAVs, 2 BSs, and 10 channels.}
	\label{fig:11}       % Give a unique label
\end{figure*}

In Fig. \ref{fig:11}, we evaluate the performance of the proposed algorithm under varying numbers of users. In the small-scale scenario, the average reward is improved by 21.74\%, 20.69\%, and 41.18\% compared to the benchmarks for the cases with 5, 7, and 9 users, respectively. In the medium-scale scenario, the proposed algorithm achieves performance gains of 14.52\%, 21.11\%, and 33.59\% for 10, 15, and 20 users, respectively. These results confirm the scalability, robustness, and adaptability of the proposed algorithm in handling varying user densities, maintaining high performance under both light and heavy network loads.
\begin{figure*}[htbp]
    \centering
    \captionsetup{justification=justified, font=small}
	\includegraphics[scale=0.275]{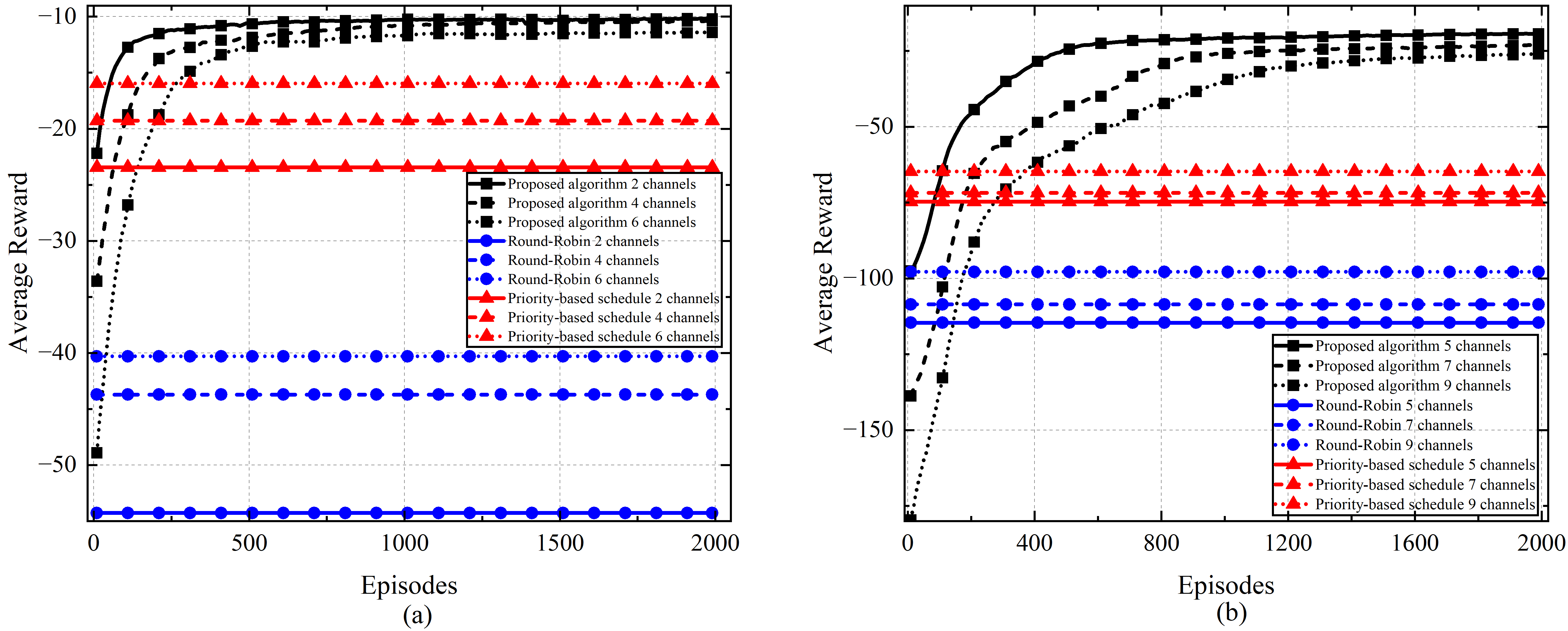}
    \caption{Performance comparison across different algorithms with varying numbers of channels. (a) Medium-scale scenario with 1 satellite, 2 UAVs, 2 BSs, and 5 users; (b) Large-scale scenario with 1 satellite, 3 UAVs, 4 BSs, and 10 users.}

	\label{fig:12}       % Give a unique label
\end{figure*}
\begin{figure*}[htbp]
    \centering
	\includegraphics[scale=0.27]{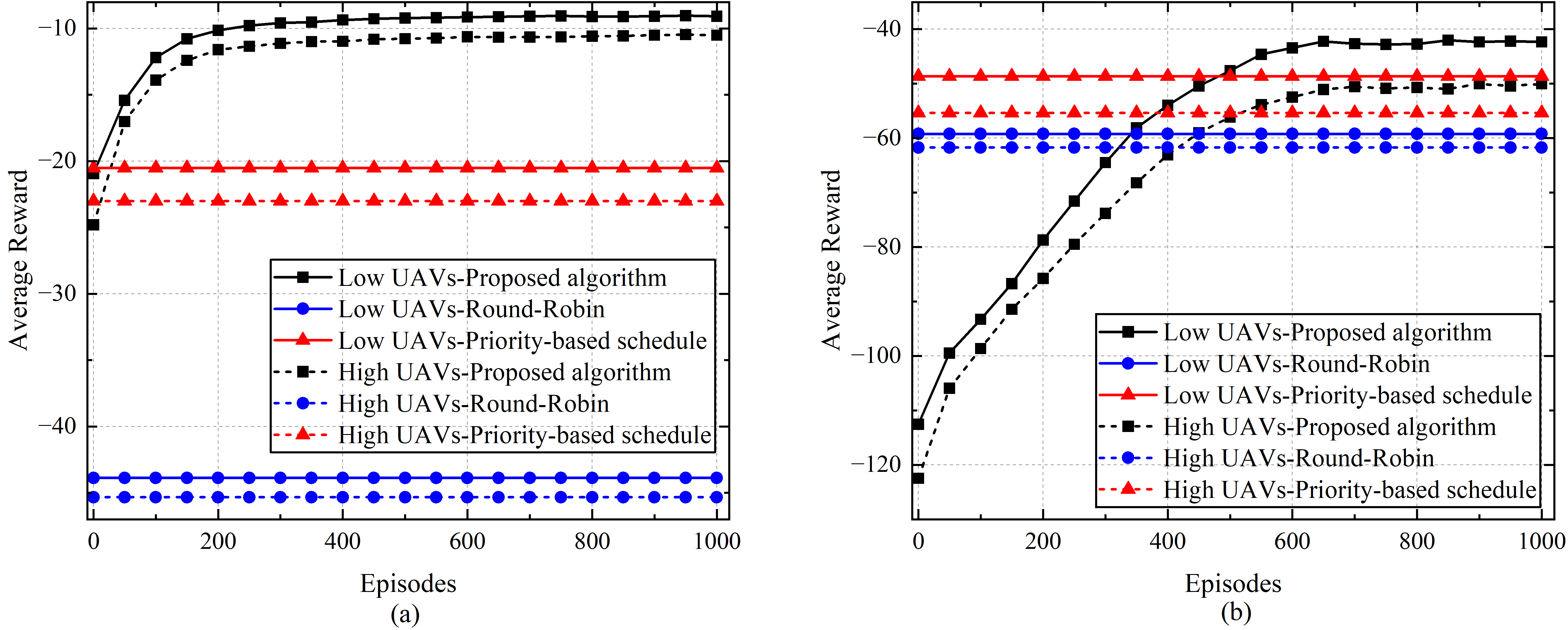}
    \captionsetup{justification=justified, font=small}
	\caption{Performance comparison across different algorithms with varying UAV heights. Low UAVs with 1 frame delay and high UAVs with 5 frames delay. (a) Medium-scale scenario with 1 satellite, 2 UAVs, 2 BSs, 3 channels, and 5 users; (b) Large-scale scenario with 1 satellite, 4 UAVs, 4 BSs, 7 channels, and 10 users.}
	\label{fig:13}       % Give a unique label
\end{figure*}

In Fig. \ref{fig:12}, we evaluate the performance of the proposed algorithm under scenarios with varying numbers of available channels. As shown in Fig. \ref{fig:12}(a), the proposed algorithm consistently outperforms the benchmarks, achieving average reward improvements of 56.67\%, 46.07\%, and 28.61\% over the priority-based schedule for 2, 4, and 6 channels, respectively. Similarly, Fig. \ref{fig:12}(b) indicates that the proposed algorithm achieves even greater performance gains, with improvements of 73.94\%, 67.91\%, and 60.10\% for 5, 7, and 9 channels, respectively. These results highlight the adaptability of the proposed algorithm to varying channel availability and its effectiveness in bandwidth-constrained environments.

In Fig. \ref{fig:13}, we evaluate the impact of UAV altitude on the performance of the proposed algorithm across different network scales. In the medium-scale scenario, the average reward is improved by 55.72\% and 54.32\% in the low and high UAV configurations, respectively. In the large-scale scenario, the proposed algorithm maintains robust performance, achieving 12.91\% and 9.68\% improvement over the benchmarks under the same low and high altitude settings. These results highlight the algorithm’s robustness to varying delay conditions and its adaptability to heterogeneous network environments.

\section{Conclusion}
\label{sec:conclusion}
In this work, we investigate the task scheduling problem in SAGUIN. We introduce the concept of the ripple effect to characterize spatiotemporal interferences in SAGUIN, and formulate its mathematical expression. With the ripple effect interference model, we formulate the task scheduling problem for SAGUIN as an MDP. To mitigate the high dimensionality challenge, the original MDP is reformulated as a Markov game. Furthermore, to handle time-varying constraints induced by ripple effect and partial observation issues, we develop a modified MAPPO algorithm. Extensive simulation results demonstrate that the proposed algorithm achieves notable improvements in AoI, energy efficiency, and overall system stability compared to conventional benchmarks. Moreover, the algorithm exhibits strong robustness and scalability across various network scales, coverage conditions, and resource constraints, highlighting its effectiveness and practicality for task scheduling in SAGUIN.

\bibliographystyle{IEEEtran}
\bibliography{IEEEabrv,main}

\begin{IEEEbiography}
[{\includegraphics[width=1in,height=1.25in,clip,keepaspectratio]{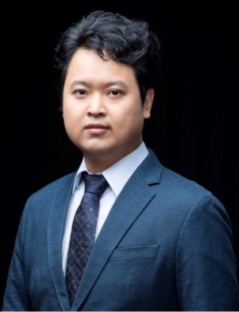}}]{Chuan Huang} (Member, IEEE) received the Ph.D. degree in Electrical Engineering from Texas A\&M University, College Station, USA, in 2012. From August 2012 to July 2014, he was a Research Associate with Princeton University and a Research Assistant Professor with Arizona State University, Tempe, USA. He is currently an Associate Professor with the Chinese University of Hong Kong, Shenzhen, China. His current research interests include wireless communications and signal processing. He served as a Symposium Chair for IEEE GLOBECOM 2019 and IEEE ICCC 2019 and 2020. He has been serving as an Editor for the IEEE Transactions on Wireless Communications, IEEE Access, the Journal of Communications and Information Networks, and the IEEE Wireless Communications Letters.
\end{IEEEbiography}

\begin{IEEEbiography}
[{\includegraphics[width=1in,height=1.25in,clip,keepaspectratio]{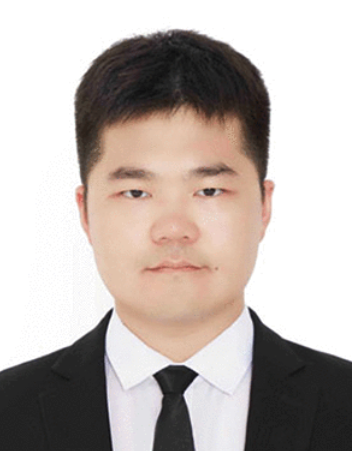}}]{Ran Li} (Member, IEEE) received the B.E. degree in communication engineering from the University of Electronic Science and Technology of China (UESTC), Chengdu, China, in 2017, and the PhD degree in computer and information engineering from the Chinese University of Hong Kong, Shenzhen, China, in 2023. He is currently a research associate with the Chinese University of Hong Kong, Hong Kong. His current research interests include reinforcement learning and near field communication.
\end{IEEEbiography}

\begin{IEEEbiography}
[{\includegraphics[width=1in,height=1.25in,clip,keepaspectratio]{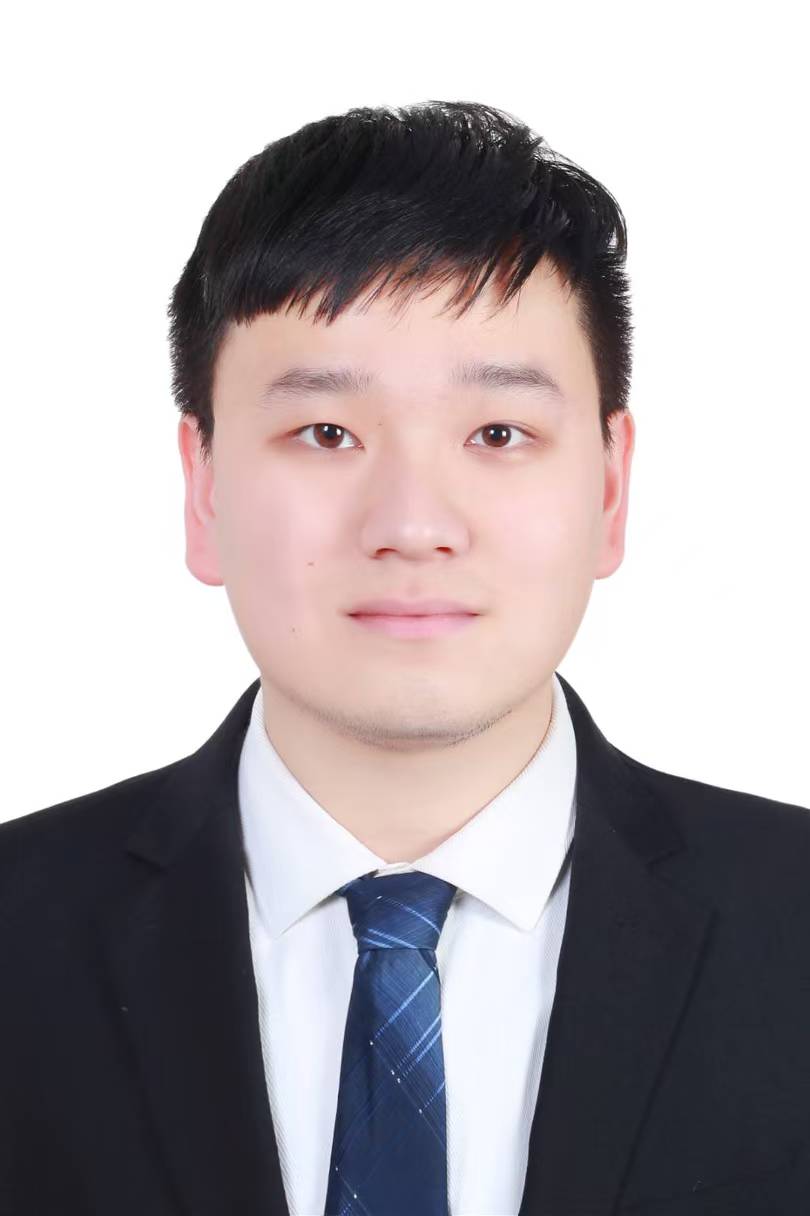}}]{Jiachen Wang} received the B.E. degree in Communication Engineering from Nanjing University of Aeronautics and Astronautics (NUAA), Nanjing, China, in 2020, and the M.E. degree in Electronic and Information Engineering from Shaanxi Normal University, Xi'an, China, in 2023. He is currently pursuing the Ph.D. degree with the School of Science and Engineering (SSE), The Chinese University of Hong Kong, Shenzhen, China. His current research interests include reinforcement learning and task scheduling.
\end{IEEEbiography}

\end{document}